\documentclass[twocolumn,showpacs,nofootinbib]{revtex4}

\usepackage{epsf}
\usepackage{graphicx}% Include figure files
\usepackage{dcolumn}% Align table columns on decimal point
\usepackage{longtable}

\newcommand{\BE}{\begin{equation}}
\newcommand{\EE}{\end{equation}}
\newcommand{\BA}{\begin{eqnarray}}
\newcommand{\EA}{\end{eqnarray}}

\def\be{\begin{equation}}
\def\ee{\end{equation}}
\def\bea{\begin{eqnarray}}
\def\eea{\end{eqnarray}}

\begin{document}
\input epsf

\draft

\renewcommand{\topfraction}{0.8}
\preprint{astro-ph/0308007, \today}

\title{\Large\bf
Observations of cluster substructure using weakly lensed sextupole
moments }

 \author{\bf John Irwin and Marina Shmakova}
\affiliation{ Stanford Linear Accelerator Center, Stanford
University, Stanford CA 94309, USA}

{\begin{abstract} Since dark matter clusters and groups may have
substructure, we have examined the sextupole content of Hubble
images looking for a curvature signature in background galaxies
that would arise from galaxy-galaxy lensing. We describe
techniques for extracting and analyzing sextupole and higher
weakly lensed moments. Indications of substructure, via spatial
clumping of curved background galaxies, were observed in the image
of CL0024 and then surprisingly in both Hubble deep fields. We
estimate the dark cluster masses in the deep field. Alternatives
to a lensing hypothesis appear improbable, but better
statistics~will be~required to exclude them conclusively.
Observation of sextupole moments would then provide a means to
measure dark matter structure on smaller length scales than
heretofore.
\end{abstract}}

% \vskip2pc]
\maketitle

%%%%%%%%%%%%%%%%%%%%%%%%% Introduction %%%%%%%%%%%%%%%%%%%%%%%%%%%%%%%%%%%%%%%%%%%%

\section{Introduction}

The observations of the Supernova Cosmology Project and the High-z
Supernova Search team \cite{supernova, Riess:1998cb, Tonry:2003zg}
as well as the Cosmic Microwave Background observation
\cite{Bond}, later confirmed by the Wilkinson Microwave Anisotropy
Probe (WMAP) \cite{Spergel:2003cb}, suggest the possibility of an
accelerating expansion of the Universe, which could mean the
presence of a positive cosmological constant or dark energy. These
observations give rise to questions as to the nature of this dark
energy/cosmological constant, possible predictions for the
evolution of the Universe based on this assumption, and possible
ways to improve or find new observational methods that could lead
to a better understanding of this evolution.

Weak gravitational lensing methods (for review see
\cite{Bartelmann:1999yn, Mellier:1998pk, Hoekstra:2002nf} and
references therein), that allow one to investigate the evolution
of matter clustering and the growth of large-scale structure
\cite{Wittman:2000tc, Mellier:2002vp}, are a way to probe both
dark energy and dark matter. In a sense it is a unique way to
investigate both the past and future of the universe. The large
structure growth is defined by the growth of the matter density
fluctuations predicted by inflationary cosmology and dark energy
evolution (for review see \cite{Peebles:xt}). The structure of
density fluctuations provides information about inflationary
scenarios \cite{Huterer:2000mj, Huterer:2001yu, Linder:2003dr}. On
the other hand, the matter distribution measurements give bounds
on the dark energy equation of state, allowing one to predict the
future of the universe \cite{Linde:2002gj,Kallosh:2003mt,
Kallosh:2003bq}.

The traditional weak gravitational lensing techniques
\cite{Kaiser:2000if}, despite observational superiority in
measuring the large clumps of matter such as clusters of galaxies
(visible or dark) with $10^{14} \, M_\odot$  are not sensitive to
the substructure of such clusters or smaller groups and clumps of
matter. We have investigated a possibility to use more sensitive
methods to expose such substructure and detect the presence of
smaller clumps. The development of these methods and their
application to analysis of data from the future observational
projects like the SuperNova /Acceleration Probe (SNAP)
\cite{unknown:2002dp, SNAP}, and the Large Synoptic Survey
Telescope (LSST) \cite{Tyson:2003kb} promise to expand our
knowledge of large scale structure to scales unreachable by other
techniques such as microwave background anisotropy or traditional
weak lensing. These observations could make an important
contribution to the understanding of dark matter structure as well
as possible inflationary scenarios.

Arclets are a familiar strong-lensing phenomena
\cite{Bartelmann:1999yn, Mellier:1998pk}. They are an example of
the general property that the nonlinear 1/r deflections of a light
stream passing a mass concentration will produce, relative to the
stream centroid, a full complement of moments. The curving seen in
the arclet can be understood as the correlated superposition of a
quadrupole and sextupole moment, with the length of the arclet
usually determined by the strength of the octupole moment.

To date, weak lensing has concentrated exclusively on quadrupole
moments - ellipticity \cite{Kaiser:2000if}, because usually all
other lensing-induced higher moments are smaller than the
lensing-induced quadrupole moments (that are already small
compared to background galaxy ellipticities). A typical measurable
cluster has a mass of 10$^{14}$ solar masses and a radius of 500
kpc. Since the strength of the quadrupole kick is proportional to
the mass and falls off like 1/r$^{2}$, one could get the same
quadrupole moment in a light stream positioned 5 kpc from an
object of 10$^{10}$ solar masses. In the latter example, since the
sextupole moment varies as 1/r$^{3}$, the sextupole moment becomes
100 times stronger, becoming larger than the intrinsic sextupole
moments of background galaxies. It occurred to us that if clusters
or groups contain an abundance of lower mass clumps, the light
streams passing through these groups might occasionally pass close
to a small mass clump producing an observable sextupole moment in
the image \cite{finalfocus}. In this case one can expect a
correlation between the direction of the quadrupole and sextupole
moments.

To investigate this hypothesis we considered the images of the
North and South Hubble Deep Fields (HDF)  \cite{Williams:1996ay,
Casertano:2000wy} as well as some other Hubble images of known
clusters, such as CL0024. We used SExtractor software
\cite{sextractor} to select either faint images (typically
23$<$m$<$29 \footnote{Our threshold setting, typically 10 times
the rms of sky noise floor,
 causes objects to appear dimmer than the total integrated
 luminosity magnitude.}) or used z-catalogs (when available)
 to identify distant galaxies with z$>$0.7.
 %%%%%%%%%%%%%%%%%%%%%%% CL0024 %%%%%%%%%%%%%%%%%%%%%%%
\begin{figure}[h!]
\centering\leavevmode\epsfysize= 6.2 cm
\epsfbox{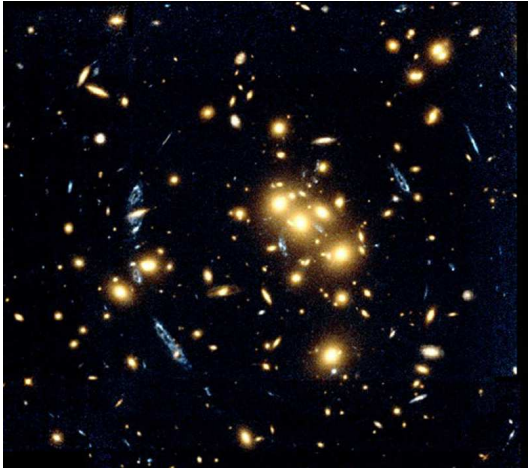}\caption[CLMath1]{ Hubble image of galaxy
cluster CL0024.} \label{CLMath1}
\end{figure}
\begin{figure}[h!]
\centering\leavevmode\epsfysize = 7 cm
\epsfbox{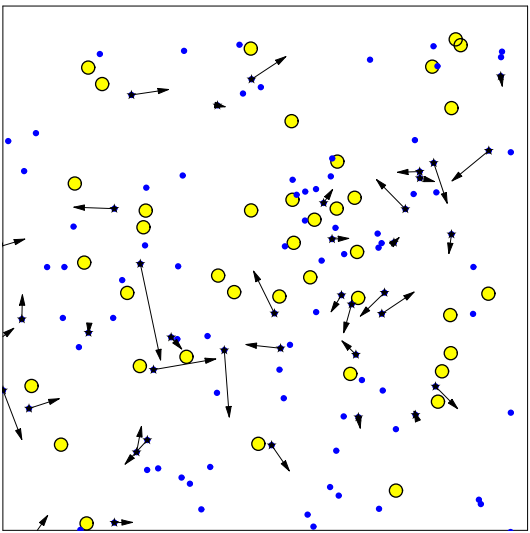} \caption[CLMath] { m-selected
background galaxies of the cluster CL0024 are indicated by blue
dots and foreground galaxies by yellow circles. The small stars
with arrows indicate the ``curved'' background galaxies with the
arrow pointing toward the scattering center and its length
proportional to the strength of the quadrupole
moment.}\label{CLMath}
\end{figure}
%%%%%%%%%%%%%%%%%%%%%%%%%%%%%%%%%%%%%%%%%%%
 The ``curved'' galaxies we sought were identified as those whose
sextupole moment was oriented so that one of its minima was
aligned within a few degrees of a quadrupole minimum. We will
refer to galaxies which have a quadrupole maximum aligned with a
sextupole maximum as `` aligned '' galaxies.

Figures  \ref{CLMath1} and \ref{CLMath} show the CL0024 cluster
and
 the location of the background galaxy sample and its ``curved'' members.
The arrows point in the direction of the scattering center and
their length is proportional to the strength of the quadrupole
moments. There is clear evidence of clumping toward the center of
this cluster. To our surprise, ``curved'' galaxies were also
clumped in both the north and south fields of the HDF survey.  The
probability for the observed clumping to
 occur by chance was about $ 1\% $  for each field.

We have carried out other tests that support the hypothesis that
the observed clumping comes from lensing: 1) ``aligned'' galaxies
were predicted and found to be clumped as strongly as curved
galaxies, 2) galaxies halfway between ``aligned'' and ''curved''
were predicted and found to have no clumping,  3)``curved''
galaxies were predicted and found to have smaller moments than all
galaxies, and 4) ``aligned'' galaxies were predicted \footnote{We
thank B.J. Bjorken for pointing this out to us.  } and found to
have moment strengths distributed as all other background
galaxies.

We have considered causes other than lensing for the spatial
clumping of curved galaxies, such as instrumental effects,
computational effects, or other physical phenomena. For example,
since the background galaxies are known to be clumped, our result
could be the consequence of the fact that, for some unknown
reason, galaxies in some groups tend to be more curved than in
other groups. Or perhaps galaxies of a certain epoch tend to be
more curved. The tests we have constructed and their implications
for each alternative hypothesis are discussed and the results
summarized in Table I.

Finally, assuming we are indeed seeing lensing, we have made an
initial attempt to deduce the mass of the objects doing the
lensing and the mass of the group in which they reside. The mass
of the group is easier to estimate than the mass of the lensing
objects themselves, in that the former depends only on the i) the
size of the observed group, ii) the fraction of background
galaxies that are lensed, and iii) the average induced quadrupole
moment. The results we present here are preliminary.

Since in some cases the footprint of the observed galaxy could
pass through a dark matter clump, we present an approach to this
general situation as well. While individual mass measurements
remain out of reach, it could be expected that modeling would
supply a basis to deduce well-defined statistical results when
analyzing larger pieces of sky.

%%%%%%%%%%%%%%%%%%%%%%%%%%%%  II %%%%%%%%%%%%%%%%%%%%%%%%%%%%%%%%%%%%%%%%%%%%%%%

\section{ Theoretical framework }

\begin{figure}[h!]
\leavevmode\epsfysize= 8 cm \epsfbox{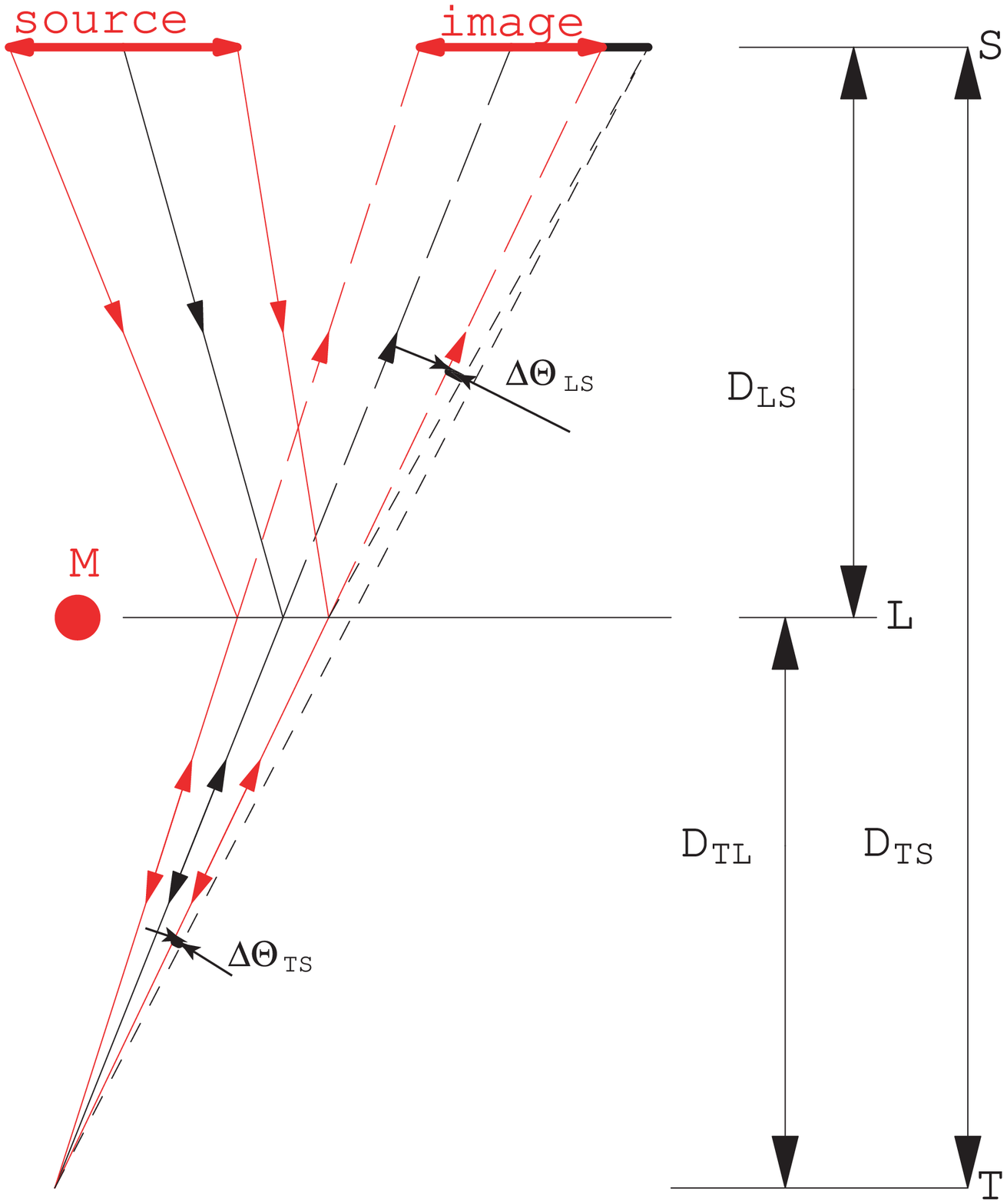}
\caption[lens] {A diagram showing 3 light rays from a source image
scattered by a single concentrated mass.  The relationship $
D_{TS} \Delta \theta_{TS} = D_{LS}  \Delta \theta_{LS}$  expresses
the observed displacement angle in terms of the deflected angle.
Note that the image in the radial direction is narrower than the
source width.} \label{lens} \leavevmode\epsfysize= 7.5 cm
\epsfbox{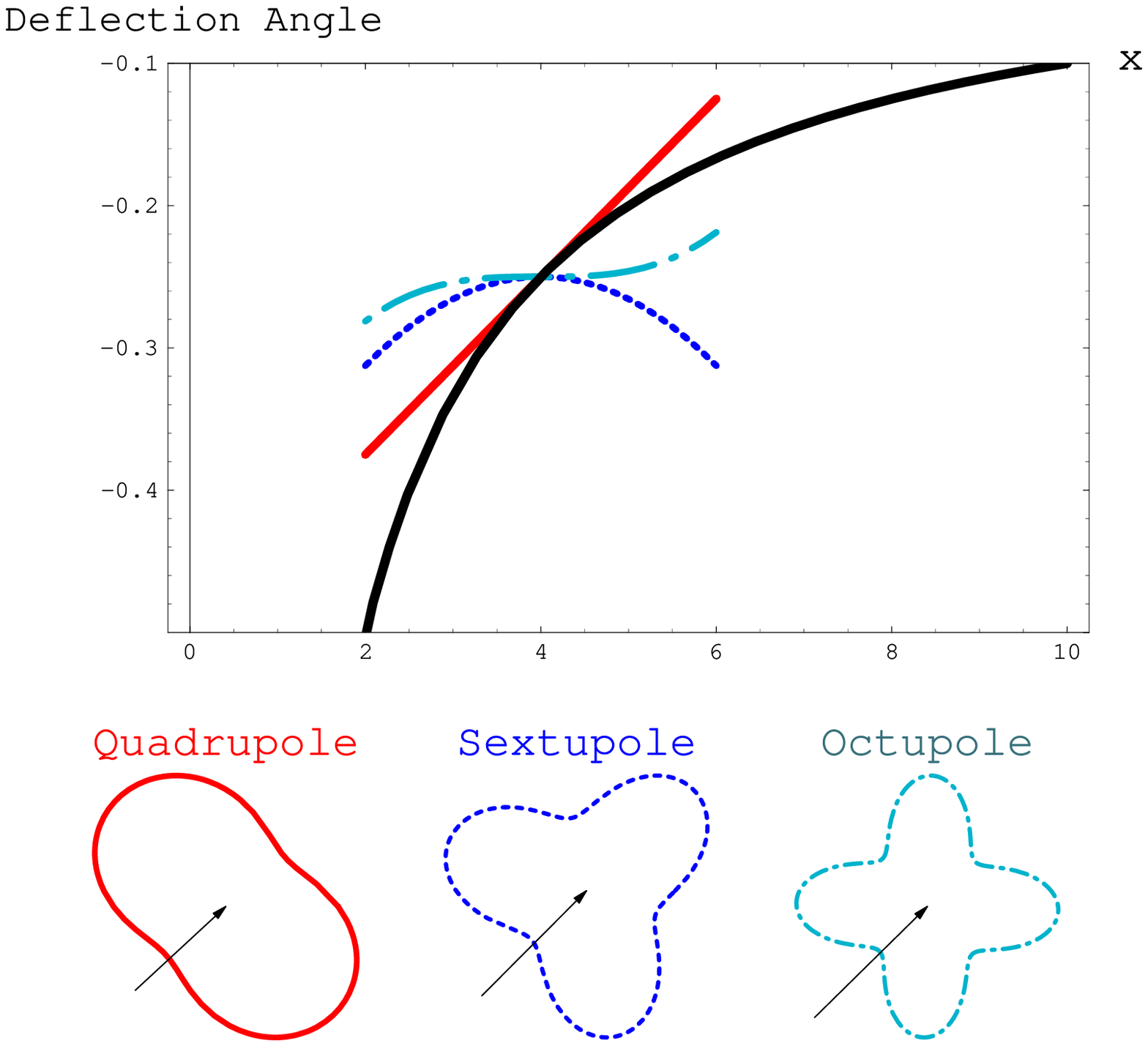} \caption[moments] {The curved line in
this graph shows the 1/r kick from a point source.  At a point $
r_0 $ this curve may be expanded in a power series (see eq.
\ref{deflangle}) yielding  linear, quadratic and cubic terms that
are shown by a tangent line, a dashed curve and a dot-dash curve,
respectively. The deformation of a constant intensity  circle in a
light beam centered at $r_0$ by each of these terms is also shown
here.  The straight lines indicate the radial direction from the
origin to the center of the light beam.  The orientation of these
images is such that there is always a deflection minimum toward
the scattering center.  The quadrupole plus sextupole yields a
slight ``banana'' shape with the radius of curvature originating
at the scattering center.} \label{moments}
\end{figure}

Fig.\ref{lens} shows rays from a distant background galaxy being
deflected by a mass concentration. The apparent image, as seen by
the telescope, is defined by an intensity function which depends
only on the angle of each ray as it enters the telescope.
Following a ray backwards, toward the apparent image, it is
deflected by mass distributions, but is known to depart somewhere
from the source galaxy. It will have a definite position and angle
at the source galaxy (measured relative to the position and angle
of the centroid ray). This map from telescope variables to source
variables will be symplectic since the light geodesics are
described by a Hamiltonian. From the point of view of the galaxy,
all rays traced backward from the telescope come from the same
point. Hence the two angles at the telescope uniquely describe the
 trajectory through space and the initial position $x_S $ and $y_S $
  (and angle ${x}'_S $ and ${y}'_S $)  at
the galaxy. So the ``backwards'' map can be written as a set of 4
functions $x_S (x_T ,y_T ),\;y_S (x_T ,y_T ),\;{x}'_S (x_T ,y_T
),\;\mbox{and }{y}'_S (x_T ,y_T )$, where ``T'' designates
``telescope'' and ``S'' designates ``source''. The coordinate
system for both the source and telescope images can be taken to be
the pixel grid on the focal plane. The position of the centroid
trajectory is taken to be the origin for both images, i.e. the
dipole kick suffered by the image as a whole is ignored. Only
under special circumstances, such as Einstein rings or
point-to-point focusing, will a ray leaving the telescope at two
different angles arrive at the same point on the source galaxy. We
will not need to consider such cases and hence will be able to
drop the two functions ${x}'_S \mbox{ and }{y}'_S $. We will be
able to assume that the determinant
\begin{equation}
\label{eq1}
\left| {{\begin{array}{*{20}c}
 {\frac{\partial x_S }{\partial x_T }} \hfill & {\frac{\partial x_S
}{\partial y_T }} \hfill \\
 {\frac{\partial y_S }{\partial x_T }} \hfill & {\frac{\partial y_S
}{\partial y_T }} \hfill \\
\end{array} }} \right|\ne 0.
\end{equation}
The two functions $x_S (x_T ,y_T )\mbox{ and}\;y_S (x_T ,y_T )$
can be combined into one complex function by defining  $w_S =x_S
+iy_S $. This complex function can be written in terms of the
variables $w_T =x_T +iy_T $ and $\bar {w}_T =x_T -iy_T $ by
substituting $x_T =\frac{1}{2}\left( {w_T +\bar {w}_T } \right)$
and $y_T =\frac{1}{2i}\left( {w_T -\bar {w}_T } \right)$. The map
equations can then be written as the single function $w_S (w_{T,}
\bar {w}_T ).$  Since the transverse width of the light stream
will be small compared to characteristic dimensions of the
variations of the mass distributions, we may expand this function
in a power series about the stream centroid:
\begin{equation}
\label{eq2}
w_S (w_T ,\bar {w}_T )=w_T +\sum\limits_{n,m=0}^\infty {a_{nm} w_T^n } \bar
{w}_T^m .
\end{equation}

The $1+a_{10} $ combination of terms is a rotation and scaling,
the $a_{01} $ term is a quadrupolar distortion, the $a_{02} $ term
is a sextupolar distortion, the $a_{03} $ term is an octupolar
distortion, the $a_{20} $ term is a cardioid-like distortion, and
the $a_{11} $ term is an $r^2$-dependent translation of circles,
and so on. We will be concerned with the terms  $a_{01} $, $a_{02}
$, $a_{03} $, and $a_{20} $ and refer to them more simply by the
letters, $a,\;\tilde {b},\;\tilde {c},\mbox{ and }\tilde {\bar
{d}}$ respectively. We have introduced the tilde symbol ``$\sim
$'' to alert the reader to the fact that these coefficients have
dimensions, and to distinguish them from dimensionless partners we
will introduce later. For a map arising from a single kick
$a_{01}$ is necessarily real and $a_{11} =2\bar {a}_{20} =2\tilde
{d}.$

Since the largest mass distribution size ($\sim $500 kpc) is still
small compared to typical path lengths ($\sim $1000 Mpc), and
since the light-deflection angles are small ($<$ 10$^{-4}$
radians) and can be calculated by multiplying the deflection
angles of a non-relativistic particle by 2, one can integrate the
transverse component of the 1/r$^{2}$ force from a point
distribution along a straight path to get the deflection angle
 \BA && \Delta {r}'  =
-\frac{4MG}{r}= - \frac{4MG}{r_0}\left( \frac{1}{1 +
\frac{r - r_0}{r_0}}\right )\label{deflangle}  \\
& \sim &   \frac{4MG}{r_0}\left( -1  + \frac{\delta r}{r_0} -
\left( \frac{\delta r}{r_0} \right )^2 + \left( \frac{\delta
r}{r_0} \right )^3 - \cdots \right ),\nonumber \EA where $\delta
r=r - r_0.$

The potential for this kick is $2\Phi _\delta =4MG\,Ln\left[ r
\right]$, which is the Green's function for the 2 dimensional
Laplace equation, $\nabla ^2\Phi =4\pi G\rho $, where $\rho $ is
the 2D (longitudinally-integrated) density function for the mass
distribution.

The usefulness of the complex variables originates in part from
the fact that the solution to Laplace's equation in empty space
can be written as the real part of an analytic function. For the
point mass source (which would be the same as outside a
symmetrical distribution) one may write  $2\Phi _\delta
=4MG\;\,\Re \left( \,Ln\left[ {x+iy} \right] \right) $. Expanding
about the centroid ray located at some  $w_0 =x_0 +iy_0$ results
in (a constant is dropped)
\begin{equation}
\label{eq3} 2\Phi _\delta =-4MG\;\; \Re \left[ \,\sum\limits_{n=1}
{\frac {1}{n} \left( {-\frac{\omega }{\omega _0 }} \right)}^n
 \right].
\end{equation}
It is useful to introduce derivative operators $\frac{\partial
}{\partial w}\equiv \frac{1}{2}\left[ {\frac{\partial }{\partial
x}-i\frac{\partial }{\partial y}} \right]\mbox{ and
}\frac{\partial }{\partial \bar {w}}\equiv \frac{1}{2}\left[
{\frac{\partial }{\partial x}+i\frac{\partial }{\partial y}}
\right]$, which have the property that $\frac{\partial w
}{\partial w}=\frac{\partial \bar {w} }{\partial  \bar {w}} =1$
and $\frac{\partial \bar {w} }{\partial w}=\frac{\partial
w}{\partial \bar {w}}=0$. In terms of these operators the kick
from the potential can be written as (the constant dipole kick is
dropped)
\begin{equation}
\label{eq4} \Delta {w}^{\prime}_\delta = -2\frac{\partial \left(
{2\Phi_\delta } \right )}{\partial \bar {w}} =-\frac{4MG}{\bar
{w}_0 }\;\,\sum\limits_{n=2} {\left( {-\frac{\bar {w}}{\bar {w}_0
}} \right)} ^{n-1}.
\end{equation}
The first three terms of this expression are the quadrupole,
sextupole and octupole, respectively (see Fig.\ref{moments}).
%%%%%%%%%%%%%%%%%% fig 2 and 3 for lensing and deflection %%%%%%%%%%%%%%%%%
%%%%%%%%%%%%%%%%%%%%%%%%%%%%%%%%%%%%%%%%%%%%%%%%%%%%%%%%%%%%%%%
The geometry of Fig.\ref{lens} shows that $\Delta w_S =D_{LS}
\;\Delta {w}'(w,\bar {w})$, where $D_{LS} $ is the
angular-diameter distance between the source and the lens. And if
one wishes to obtain the map as a function of the telescope
variables rather than the lensing plane variables one may
substitute$\mbox{ }w=\frac{D_{TL} }{D_{TS} }w_T $.

It follows that the coefficients $a,\;\tilde {b},\;\mbox{and
}\tilde {c}$ of this point-source map are given by
 \BA \label{eq5}
  a_\delta & = &D_{LS} \frac{D_{TL}}{D_{TS}
}\frac{4MG}{\bar {w}_0^2 },\cr
 \tilde {b}_\delta & = & -D_{LS} \left(
{\frac{D_{TL} }{D_{TS} }} \right)^2\frac{4MG}{\bar {w}_0^3 },\cr
\tilde {c}_\delta & = & D_{LS} \left( {\frac{D_{TL} }{D_{TS} }}
\right)^3\frac{4MG}{\bar {w}_0^4 }. \EA The magnitudes of interest
will be the rms value of the dimensionless quantities $a_\delta
,\;b=\tilde {b}_\delta \sigma ,\mbox{ and }c=\tilde {c}_\delta
\sigma ^2$, where \textit{$\sigma $}  is the rms size of the
image. Each of these coefficients decreases in magnitude from the
previous by $r_G /r_0 $, where $r_G \equiv \frac{D_{TL} }{D_{TS}
}\sigma $ is the footprint radius of the background galaxy at the
lensing mass plane and $r_{0}$ is the distance from the center of
the footprint to the center of the lensing mass.

For a potential derived from a solution of Laplace's equation for
a general mass distribution, one can expand the potential about
the centroid ray to get its local power series expression. The map
obtained from the derivative of this power series potential will
necessarily be of the form of equation (\ref{eq2}). Since the
potential is a real function, its expansion in terms of $w$ and
$\bar {w}$ will have some constraints:

\begin{enumerate}
\item the coefficients of the linear $w$ and $\bar {w}$ terms must
be the complex conjugate of one another, so there is one complex
parameter, which is the magnitude and direction of the dipole
kick; \item the coefficients of the $w^{2 }$and $\bar {w}^2$  must
be the complex conjugate of one another, and represent the
magnitude and orientation of the quadrupole kick; \item the
coefficient of $w\bar {w}$ must be real (the coefficient of this
term must be proportional to $\left. {\frac{\partial }{\partial
w}\frac{\partial }{\partial \bar {w}}\Phi } \right|_0
=\frac{1}{4}\left. {\nabla ^2\Phi } \right|_0 =\pi G\left. \rho
\right|_0 $ and hence is zero except when the light path passes
through a distribution ) and represents a magnification; \item the
coefficients of the $w^{3 }$ and $\bar {w}^3$  must be the complex
conjugate of one another, and represent the magnitude and
orientation of the sextupole kick, \item the coefficients of the
$w^2\bar {w}$  and $w \bar {w}^2$  must be the complex conjugate
of one another, and represent the magnitude and orientation a
third order term which is also necessarily zero except when the
light path passes through a distribution. In this last case, the
kick can be written in the form $\Delta w=\bar {\tilde {d}}{\kern
1pt}w_T^{2}+2\tilde {d}{\kern 1pt}w_T\bar {w_T}$.\footnote{ The
complex notation is awkward in this case. For real d the potential
function is $\Phi \sim - d r^{3} \cos \theta $, from which $\Delta
r^{\prime} \sim  + 3 d r^{2}  \cos \theta $ and $ r \Delta
\theta^{\prime} \sim  - d r ^{2}  \sin \theta $.} $\tilde {d}$
will be proportional to the first derivatives of \textit{$\rho $},
in fact $$\tilde {d}=-2\pi G\,D_{LS} \left(
\frac{D_{TL}}{D_{TS}}\right)^2 \left. {\frac{\partial \rho
}{\partial \bar {w}}} \right|_0. $$
\end{enumerate}
 The final element we will need in a minimum
theoretical framework is a method to deduce the coefficients of
the map from the image. This process begins by noting the surface
brightness relationship $i_S (x_S ,y_S )\,dx_S dy_S =i_T (x_T ,y_T
)\,dx_T dy_T $, expressing the fact that if the area element is
transformed according to the map, the number of photons leaving
the source in that area will be the number observed in the image.
Defining the (unknown) moments of the source through $M_{nm}^S
\equiv \int {w_S^n } \bar {w}_S^m \,i_S (x_S ,y_S )dx_S dy_S $,
and transforming to telescope variables yields \BA \label{eq6}
&& M_{nm}^S   =  \\
& & \int {w_S^n (w_T ,\bar {w}_T )} \bar {w}_S^m (w_T ,\bar {w}_T
)i_T (x_T ,y_T )dx_T dy_T.\nonumber
 \EA
This equation can be used to
determine an expression for the quadrupole map coefficient:
\begin{equation}
\label{eq7}
\begin{array}{c}
 M_{20}^S
 =\int {\left( {w_T +a\bar {w}_T } \right)^2\,} i_T (x_T ,y_T )dx_T dy_T
\\
\nonumber \\
 =M_{20}^T +2aM_{11}^T +a^2M_{02}^T.
 \end{array}
\end{equation}
The moments on the right hand side of this equation can be deduced
from the telescope image. One obtains a quadratic equation for
$a$: one can assume $M_{20}^S =0,$ or one can find a statistical
ensemble for $a$ in terms of an assumed statistical ensemble for
 $M_{20}^S $. Note that $w\bar {w}=r^2\mbox{ so }M_{11}^T $ is the
mean square radius of the telescope image. We normalize surface
brightness functions so that $\int {i_T (x_T ,y_T )\,dx_T dy_T =1}
$.

Solving for $a$,
\begin{equation}
\label{eq8} a=-\frac{\Delta M_{20} }{2M_{11}^T }-a^2\frac{\bar
{M}_{20}^T }{2M_{11}^T },
\end{equation}
where we introduced the notation: $$ \Delta M_{nm} =M_{nm}^T -
M_{nm}^S. $$
 Since $\left[ {\bar {M}_{20}^T
/2M_{11}^T } \right]_{rms} \approx 0.15$, the last term in
(\ref{eq8}) is typically only a 2{\%} correction. Though equation
(\ref{eq7}) may be solved exactly to get
\begin{equation}
\label{eq9} a=-\frac{\Delta M_{20} }{2M_{11}^T }\frac{2}{1+\sqrt
{1-\frac{\Delta M_{20} }{M_{11}^T }\frac{\bar {M}_{20}^T
}{M_{11}^T }} }\quad ,
\end{equation}
we will be content to approximate the 2$^{nd}$ factor by unity.

To obtain an equation for the sextupole moment consider
\begin{equation}
\label{eq10}
\begin{array}{c}
 M_{30}^S =\int {\left( {w_T +a\bar {w}_T +\tilde {b}\bar {w}_T^2 }
\right)^3\,} \,i_T (x_T ,y_T )dx_T dy_T \\
 =M_{30}^T +3\tilde {b}\,M_{22}^T +3a\,M_{21}^T +\ldots \\
 \end{array}\quad .
\end{equation}
whence
\begin{equation}
\label{eq11} \tilde {b}\,\approx -\frac{\Delta M_{30}}{3M_{22}^T
}-\frac{a\,M_{21}^T }{M_{22}^T }\quad .
\end{equation}
Using equation (\ref{eq8}) for $a $  and introducing the
\textit{rms }radius of the galaxy to obtain a dimensionless
quantity we have
\begin{equation}
\label{eq12} b\equiv \tilde {b}\sigma \,\approx \left[ {-\Delta
M_{30} + \frac{3}{2}\frac{\Delta M_{20}}{M_{11}^T }M_{21}^T }
\right]\frac{\sqrt {M_{11}^T } }{3M_{22}^T }.
\end{equation}
where $\sigma = \sqrt {M_{11}^T } $.  Here we note the interesting
fact that even with $b=0 $ the quadrupole term in the map can
induce a change in the sextupole moment $ \Delta M_{30} $ if
$M_{21}^T \ne 0$.

An $M_{21}^T $ could originate in the background galaxy or be
generated by the 3$^{rd}$ order kick proportional to the
derivative of $\rho .$ To examine the latter possibility we look
at
$$M_{21}^S =\int {\left( {w_T +\Delta w} \right)^2\left( {\bar
{w}_T +\Delta \bar {w}} \right)\,} \,i_T (x_T ,y_T )dx_T dy_T ,$$
with $ \Delta w=a\bar {w}_T +\tilde {b}\bar {w}_T^2 +2\tilde
{d}{\kern 1pt}w_T \bar {w}_T +\bar {\tilde {d}}{\kern 1pt}w_T ^2$.
The result is the equation
\begin{equation}
\label{eq13} M_{21}^S =M_{21}^T +2 a \bar {M}_{21}^T + \bar {a}
M_{30}^T + 5 \tilde {d}M_{22}^T+\ldots
\end{equation}
\BA  d &= &\tilde {d}\sigma \label{eq14} \\ \nonumber \\
 &\approx & \left[ {-\Delta
M_{21}
 +\frac{\Delta M_{20} }{M_{11}^T }\bar
{M}_{21}^T }+ \frac{\bar {\Delta M_{20}} }{2M_{11}^T }{M}_{30}^T
\right] \frac{\sqrt {M_{11}^T } }{5M_{22}^T }, \nonumber
 \EA
 $\tilde {d}$ has
the direction opposite to the density gradient and in typical
cases would be expected to be along the axis of the minima of the
induced quadrupole moment.

Finally, since the telescope image has been blurred by the
point-spread function, one needs to know the moments of the
point-spread function, and if possible correct its systematic
effects on the properties of the measured image. If the complete
(no thresholding) image was available one could compute (``\^{}''
indicates convolved with the PSF, and $\Delta x_T \equiv x_T
-{x}'_T $ etc.)
 \BA  \label{eq15} \hat {M}_{nm}^T & = &\int {w_T^n \bar {w}_T^m
\,\hat {i}_T (x_T ,y_T )dx_T dy_T }  \\
 & = &\int {\int {w_T^n \bar {w}_T^m \,p(\vec {r}_T -\vec {{r}'}_T )\,i_T ({\vec
{r}}'_T )dx_T dy_T d{x}'_T d{y}'_T } } \nonumber \\
 & = & \int { \int { \left( {\Delta w_T +{w}'_T } \right)^n \left( {\Delta \bar
{w}_T +\bar {{w}'}_T } \right)^m  \times }} \nonumber \\
 & \, &
\quad \quad p(\Delta \vec {r}_T )i_T ({\vec {r}}'_T )d\Delta x_T
d\Delta y_T d{x}'_T d{y}'_T .\nonumber
 \EA
 When the binomial powers are expanded, the double
integral reduces to the sum of a product of single integrals. One
finds, for example,
\begin{equation}
\label{eq16} \hat {M}_{20}^T =M_{20}^P +2M_{10}^P M_{10}^T
+M_{20}^T =M_{20}^P +M_{20}^T.
\end{equation}
The left hand side can be determined directly. The first term on
the right hand side is the quadrupole moment of the point-spread
function, which can be determined presumably by looking at star
images. The dipole terms in the central product are both zero by
definition of the centroid. The last term is the sought after
quadrupole moment of the pure image. Note that if the point-spread
function has no quadrupole term the blurred image has the same
quadrupole moment as the original image. However the map
coefficient $a$ also involves the moment $M_{11}^T $, which is
different:
 \BA \label{eq17} \hat {M}_{11}^T  & = & M_{11}^P +M_{10}^P
M_{01}^T +M_{01}^P M_{10}^T +M_{11}^T \\
 & = & M_{11}^P +M_{11}^T . \nonumber
 \EA
This is a statement that the rms of the final image is equal to
the rms of the original image plus the rms of the point-spread
function. For completeness, for remaining moments of interest we
have \BA \label{eq18}
 \hat {M}_{30}^T &  = & M_{30}^T +M_{30}^P, \\
 \nonumber \\
 \label{eq18a } \hat {M}_{21}^T &  = & M_{21}^T
+M_{21}^P , \EA
 and
 \BA
 \hat {M}_{22}^T & = &  M_{22}^T +2M_{11}^P M_{11}^T +M_{22}^P +M_{20}^P M_{02}^T
+M_{02}^P M_{20}^T   \nonumber  \\
\nonumber \\
 &\approx & M_{22}^T +2M_{11}^P M_{11}^T +M_{22}^P .  \label{eq18a}
 \EA
In all cases, if the moments of the point-spread-function are known, the
moments of the original image can be found from the smeared image. However
this result does not strictly hold when the image is taken to be only those
pixels whose count number exceeds some threshold. We will not discuss this
problem further here.

%%%%%%%%%%%%%%%%%%%%%%%%%%% III %%%%%%%%%%%%%%%%%%%%%%%%%%%%%%%%%%%%%%%%%%%%

\section{Galaxy selection}

\begin{figure}[htbp]
\centerline{\includegraphics[width=1.50in,height=3in]{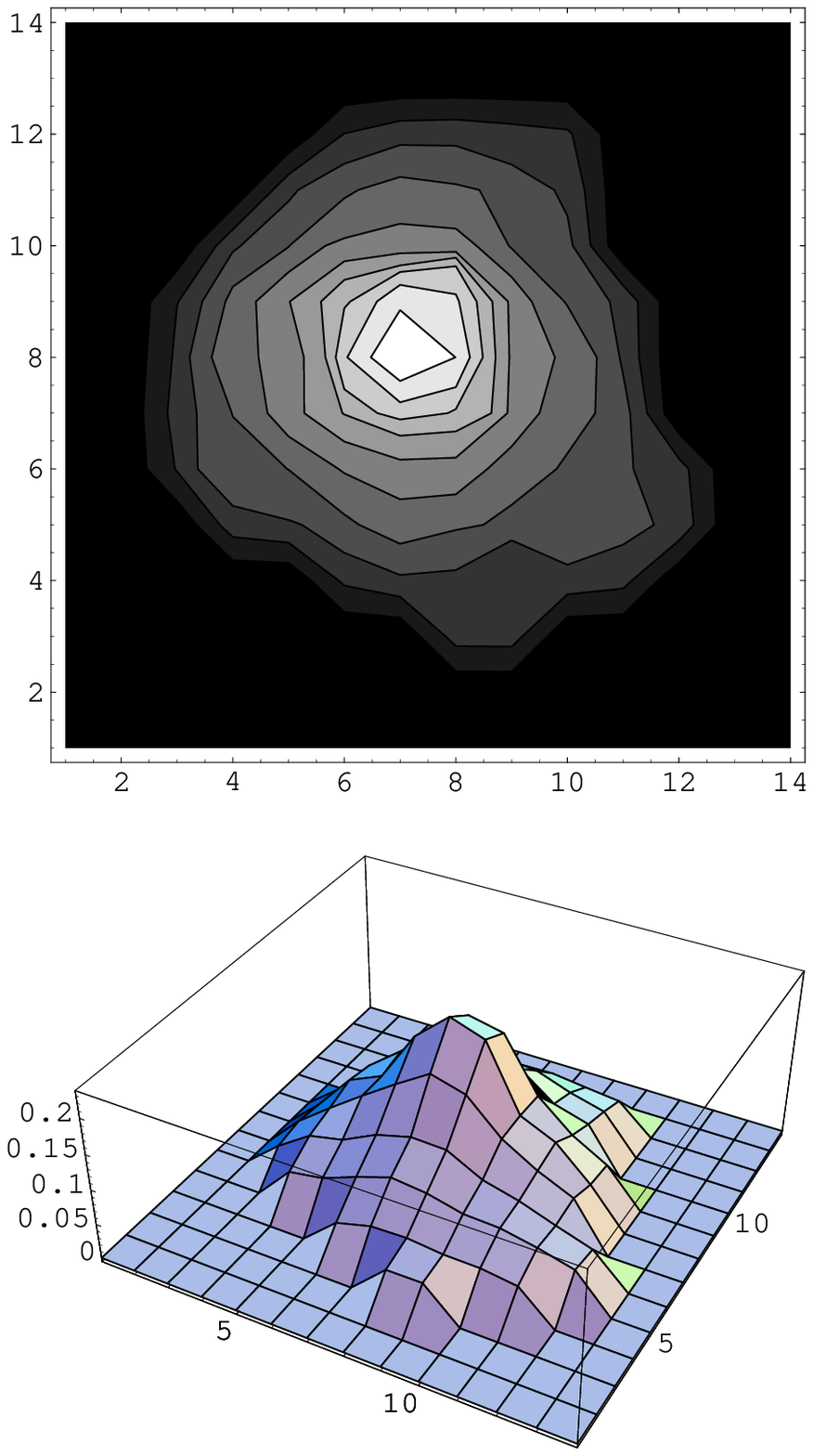}\,\,
\includegraphics[width=1.50in,height=3in]{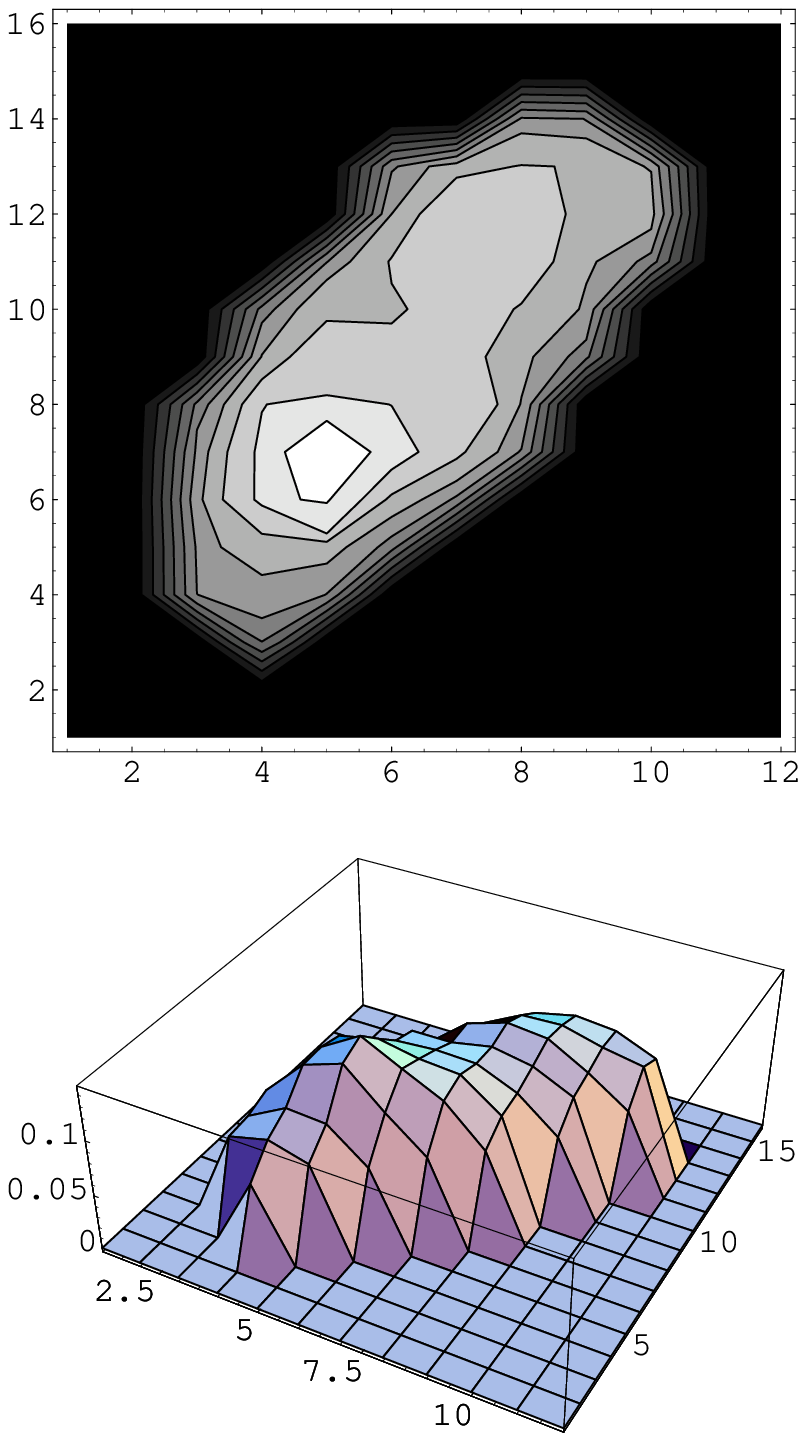}} \caption[MathG]
{Two galaxy images (contour plots and 3-D plots
of surface brightness from the Hubble north field).} \label{MathG}
\end{figure}

The software SExtractor was used to select galaxies from the
Hubble deep field and to specify which pixels to include in the
image. Galaxy images were transferred to the \textit{Mathematica}
programming environment (see Fig. \ref{MathG}). Galaxies with more
than one maxima were eliminated.

%%%%%%%%%%%%%%%%%%%%%% Mathematica galaxy %%%%%%%%%%%%%%%%%%%%%%%%%%%
%%%%%%%%%%%%%%%%%%%%%%%%%%%%%%%%%%%%%%%%%%%%%%%%%%%%%%%%%%%%%%%%%%

 Two SExtractor input parameters proved to be significant: i) the threshold setting,
 which we typically took to be 10 times the noise floor, and ii) the
 convolution matrix for pixel selection, which we typically took to be 3x3
 or a delta function. If the convolution matrix was too
broad (larger than 3x3 pixels), important pixels were dropped
resulting in a change in the computed moments. That the dropped
pixels were important was deduced from the fact that the clumping
result described below becoms less significant.

%%%%%%%%%%%%%
%%The high threshold is also necessary to observe clumping. We have
%%tried  $ 3.5 \, \sigma $  to  $ 8 \, \sigma $  and all were worse
%%than  $ 10 \, \sigma $  the sky noise floor. {\bf (***** ?????
%%****) } In fact the clumping result disappears below $ 5 \, \sigma
%%$ . Visual examination of galaxy images reveals that at lower
%%thresholds the edges are indeed less crisp, and because the
%%sextupole moment is weighted by $r^{3}$, this moment can be
%%dominated by edge effects.
%%%%%%%%%%%%%%%% change %%%%%%%%%%%%%%%%%%%%%%
The high threshold is important to observe clumping. In fact the
clumping result become less apparent below $ 5 \, \sigma $. Visual
examination of galaxy images reveals that at lower thresholds the
edges are indeed less crisp, and because the sextupole moment is
weighted by $r^{3}$, this moment can be dominated by edge effects.

Setting the threshold at $ 10 \, \sigma $  noise floor results in
images which are galaxy cores. About half of all faint galaxies
are lost in this cut. Since some fainter galaxies also have
``crisp'' cores, there may be ways to recover those by imposing a
threshold that scales with core brightness.

%%%%%%%%%%%%%%%%%%% fig 5 r rms %%%%%%%%%%%%%%%%
\begin{figure}[h!]
\leavevmode\epsfysize= 5.5 cm \epsfbox{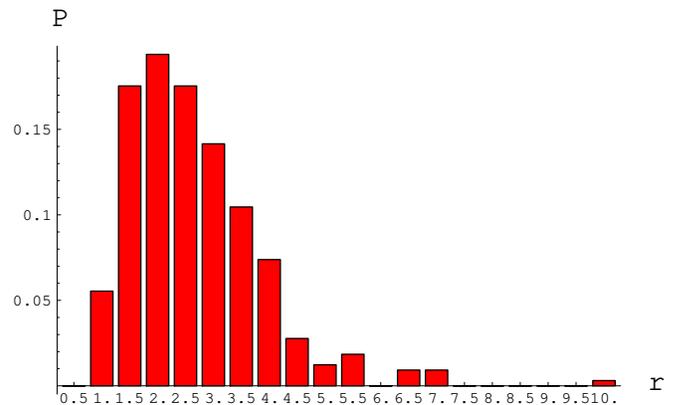} \caption[rr]
{Distribution of galaxy rms core sizes.} \label{rr}
\end{figure}
%%%%%%%%%%%%%%%%%%%%%%%%%%%%%%%%%%%%%%%%%%%%%%%%%%
Our galaxy-core images have a small rms radius of about 2 pixels.
The distribution of rms radii is shown in  Fig. \ref{rr}. The
distance across the image is typically 5 pixels. For comparison
the rms PSF radius for the Hubble deep field is 1.7 pixels.
However for high thresholds the point-spread-function spreads the
image less than indicated by its rms radius.

%%%%%%%%%%%%%%%%%%% IV %%%%%%%%%%%%%%%%%%%%%%%%%%%%%%%%%%%

\section{ Map coefficient strengths}

\begin{figure}[h!]
\leavevmode\epsfysize= 5.5 cm \epsfbox{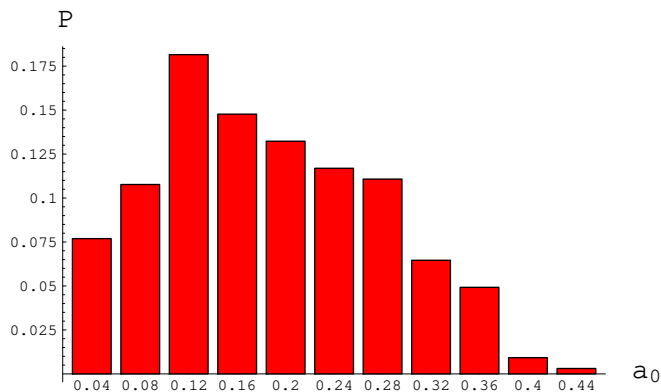} \caption[MODA] {
Distribution of the dimensionless  quadrupole moment coefficient,
$ | a_0 | $.} \label{MODA}
\end{figure}

\begin{figure}[h!]
\leavevmode\epsfysize= 5.4 cm \epsfbox{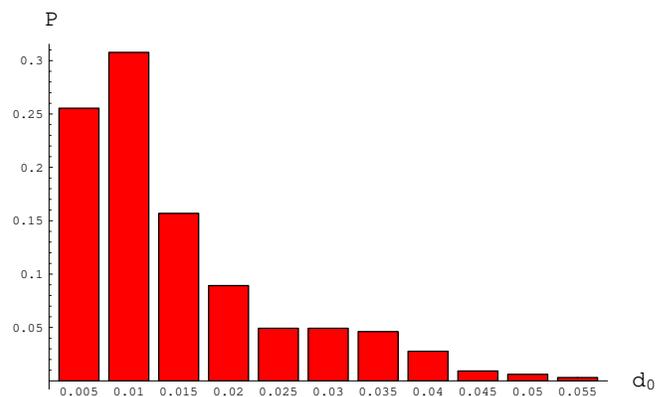} \caption[MODD] {
Distribution of the dimensionless coefficient  $ | d_0 |$.}
\label{MODD}
\end{figure}

We have carried out studies for galaxies that have been selected
by magnitude or by z-value \footnote{ We used the z-catalogs from
www.ess.sunysb.edu/astro/hdf.html and
bat.phys.unsw.edu.au/~fsoto/hdfcat.html.}. Results for all
quantities were similar. Fig. \ref{MODA} shows the measured
distribution of the quadrupole $a$ coefficient for
magnitude-selected north field galaxies with \BE \label{aa0}\hat
{a}_0 \equiv -\frac{M_{20}^T }{2\hat {M}_{11}^T }, \EE
 ($M_{20}^S $ is assumed to be zero, `` $ \hat{}
$ '' indicates smeared quantities are used). The $x$ and $y$
components of the coefficient are well fit by a Gaussian
distribution with $\hat {\sigma }_{a0} $= 0.14. The distribution
for the magnitude of $a$ is the product of two Gaussians
integrated over angle (a Rayleigh distribution) characterized by
the same $\hat {\sigma }_{a0} .$ The quadrupole moments of
selected galaxies were also calculated using a
``weighting-function'' technique. The two methods give similar
results for the $a $ coefficient. $^{ }$\footnote{ We thank David
Wittman for providing this result.}
%%%%%%%%%%%%%%%%%%%%% Fig 6 a %%%%%%%%%%%%%%%%%%%%%%%%%%
%%%%%%%%%%%%%%%%%%%%%%%%%%%% a -aw %%%%%%%%%%%%%%%%%%
%%\begin{figure}[h!]
%%\leavevmode\epsfysize= 5.4 cm \epsfbox{hsnAAW.eps}
%%\caption[MODAAW] { Distribution of the dimensionless  quadrupole
%%moment coefficient, $ | a_0 | $ versus  $ | a_W |$ calculated with
%%weighting-function. } \label{MODAAW}
%%\end{figure}
%% Fig 7
%%%%%%%%%%%%%%% fig 7 b %%%%%%%%%%%%%%%%%%%%%%%%%%%%%%%%%%%%%%%%%%%
\begin{figure}[h!]
\leavevmode\epsfysize= 5.4 cm \epsfbox{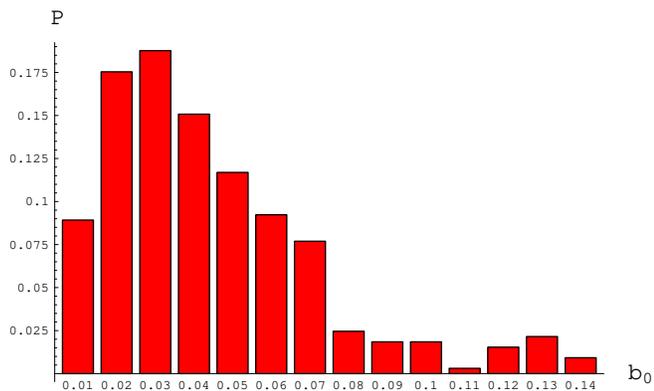} \caption[MODB] {
Distribution of the dimensionless  sextupole moment coefficient,
$| b_0 | $.} \label{MODB}
\end{figure}
%%%%%%%%%%%%%%%%%%%%%%%%%%%%%%%%%%%%%%%%%%%%%%
Fig. \ref{MODB} shows the measured distribution of the
dimensionless sextupole coefficient
\BE \label{bb0} \hat {b}_0
\equiv -\frac{M_{30}^T }{3\hat {M}_{22}^T }\sqrt {\hat {M}_{11}^T
} ,\EE ($M_{30}^S $ is assumed to be zero, the $\Delta M_{20}^T $
term is ignored) for m-selected galaxies in the north field. The
$x$ and $y$ component of the coefficient is again Gaussian with
$\hat {\sigma }_{b0} $=0.034 being a factor of 4 smaller than the
quadrupole moment $\hat {\sigma }_{a0} $.
%% Fig 7
 Fig. \ref{MODD} shows the measured distribution for the
sextupole-order $d$ coefficient \BE \label{dd0} \hat {d}_0 \equiv
-\frac{M_{21}^T }{5\hat {M}_{22}^T }\sqrt {\hat {M}_{11}^T  }, \EE
( $ M_{21}^S $ is assumed to be zero and the $ \Delta M_{30}
\mbox{ and } \Delta M_{20} $ terms are ignored ) for m-selected
galaxies in the north field.  The $x$ and $y$ component of the
coefficient are again Gaussian. $\hat {\sigma }_{d0} $=0.009, is a
factor of more than 15 smaller than the quadrupole moment $\hat
{\sigma }_{a0} $ and about a factor of 4 smaller than $\hat
{\sigma }_{b0} .$ Referring to equation (\ref{eq12}) we see that
the magnitude of the 2$^{nd}$ correction term can be estimated to
be typically a factor of 10 smaller than the leading term. In
equation (\ref{eq14}) we can estimate that the 2$^{nd}$  and
3$^{rd}$ term are typically a factor of 3 smaller than the leading
term.
%%%%%%%%%%%%%%%%%%%%% fig 8 d %%%%%%%%%%%%%%%%%%%%%%%%%%%
%%%%%%%%%%%%%%%%%%%%%%%%%%%%%%%%%%%%%%%%%%%%%
%%%%%%%%%%%%%%% fig 9 c %%%%%%%%%%%%%%%%%%%%%%%%%%%%%%%%%%%%%%%%%%
\begin{figure}[h!]
\leavevmode\epsfysize= 5.4 cm \epsfbox{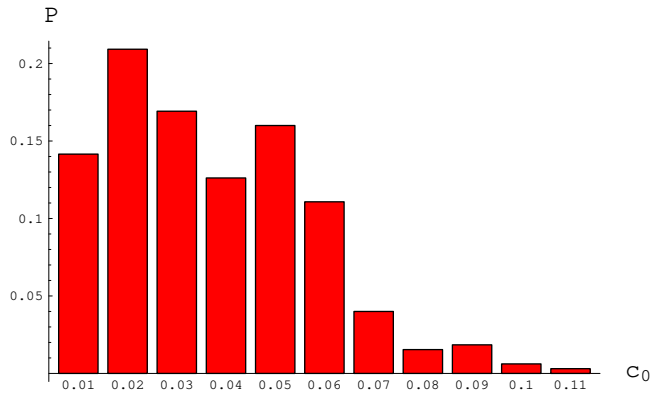} \caption[MODC] {
Distribution of the dimensionless octupole moment coefficient, $ |
c_0 | $. } \label{MODC}
\end{figure}
%%%%%%%%%%%%%%%%%%%%%%%%%%%%%%%%%%%%%%%%%%%%%%%%%%%%%%%%%%%%%%%%

Finally,  Fig. \ref{MODC} shows the distribution of the
dimensionless octupole coefficient
\BE \label{cc0}\hat {c}_0
\equiv -\frac{M_{40}^T }{4\hat {M}_{33}^T }\hat {M}_{11}^T. \EE
It
is not much smaller than the sextupole moment: $\hat {\sigma
}_{c0} $= 0.026. We note that a bi-Gaussian intensity distribution
with unequal major and minor axes will have a significant octupole
moment aligned with its quadrupole moment.

%%%%%%%%%%%%%%%%%%% V %%%%%%%%%%%%%%%%%%%%%%%%%%%%%%%%%%%%%%%%%%%%%%%%%%

\section{ Map coefficient orientation}
 %%%%%%%%%%%%%%%%%%%  fig 10 Q-S %%%%%%%%%%%%%%%%%%%%%%%%%%%%%%%
%%%%%%%%%%%%%%%%%%%%  fig 11 shapes  %%%%%%%%%%%%%%%%%%%%%
\begin{figure}[h!]
\leavevmode\epsfysize= 5.5 cm \epsfbox{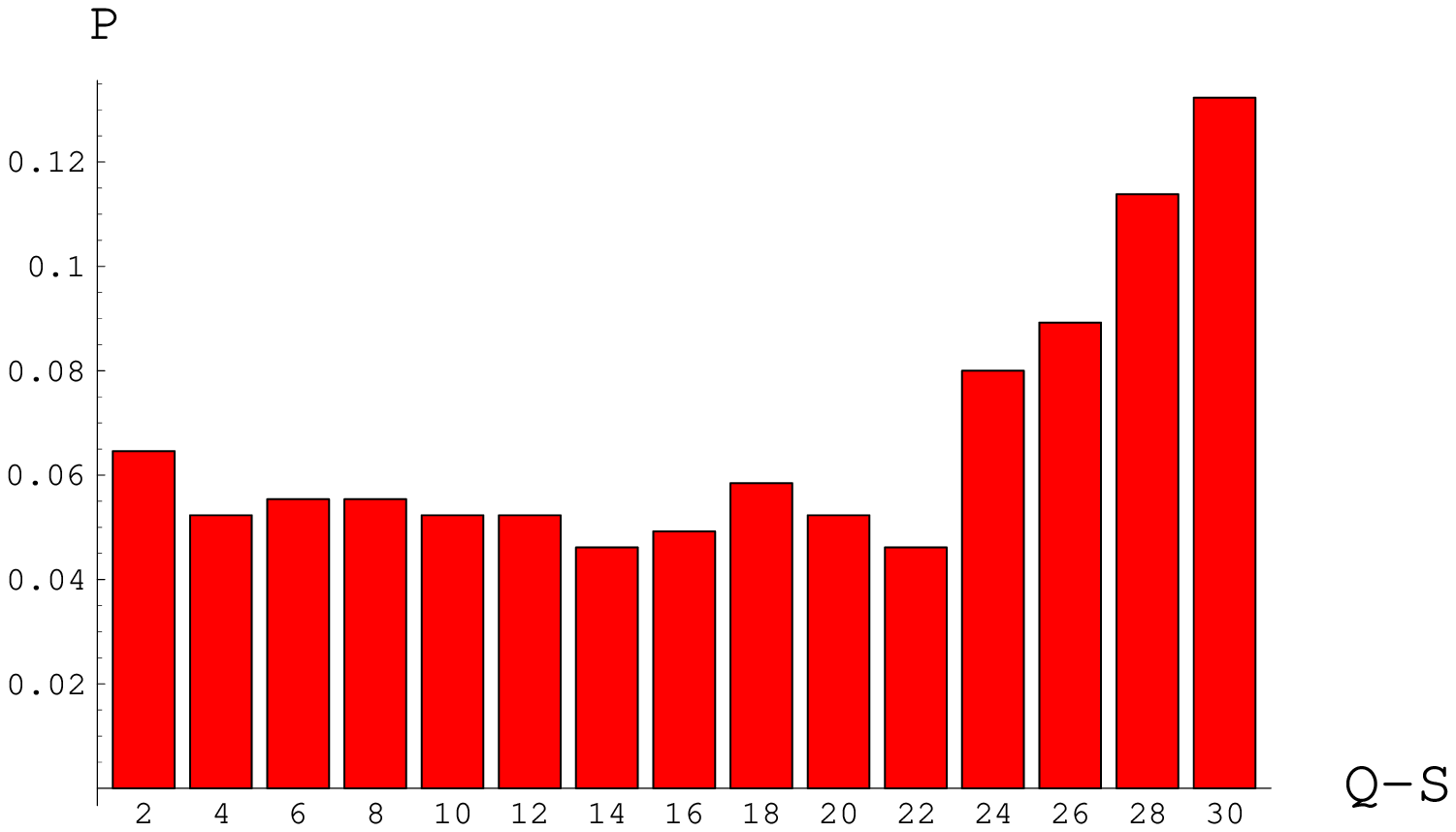}
\caption[QSCOR] {Distribution of the smallest angle btween a
quadrupole minimum and a sextupole minimum for 324 galaxies in the
north HDF. }\label{QSCOR}
\centerline{\includegraphics[width=1.3in,height=2.3in]{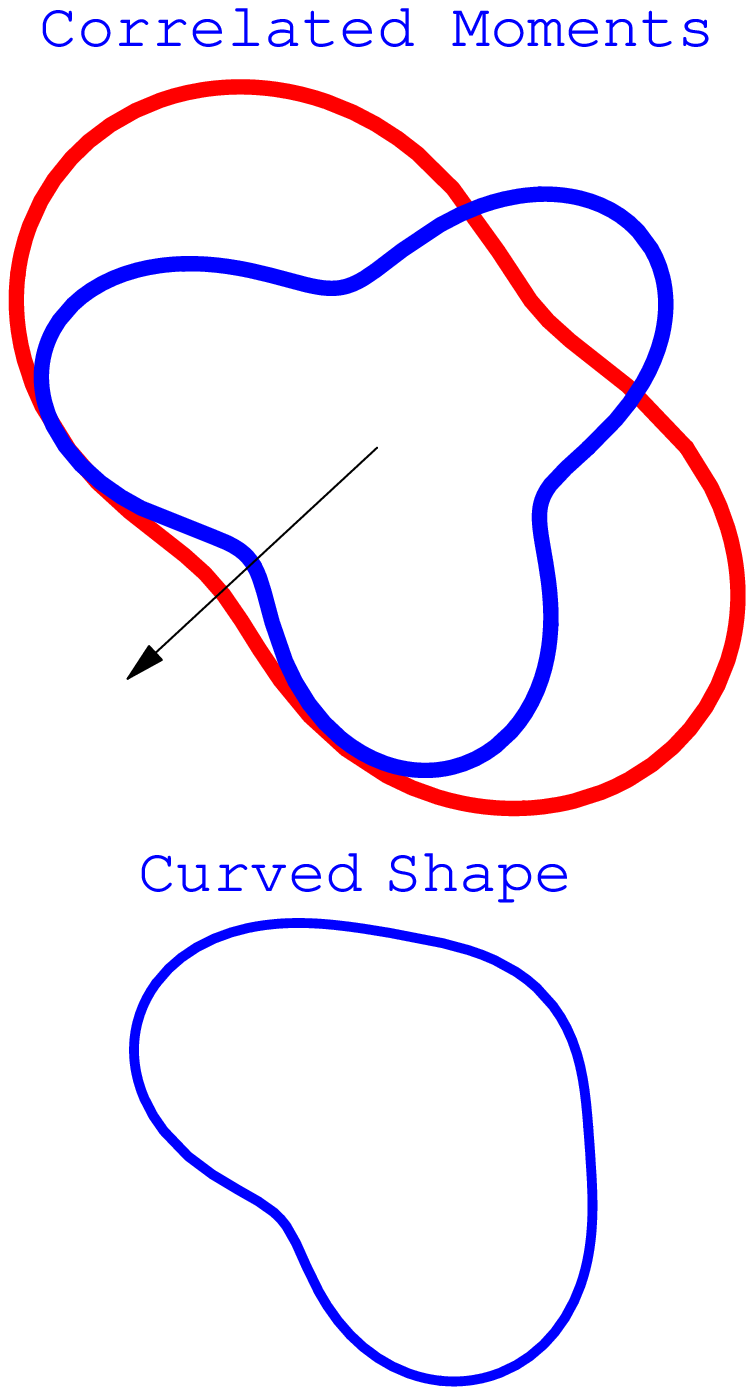}\,\,
\includegraphics[width=1.3in,height=2.3in]{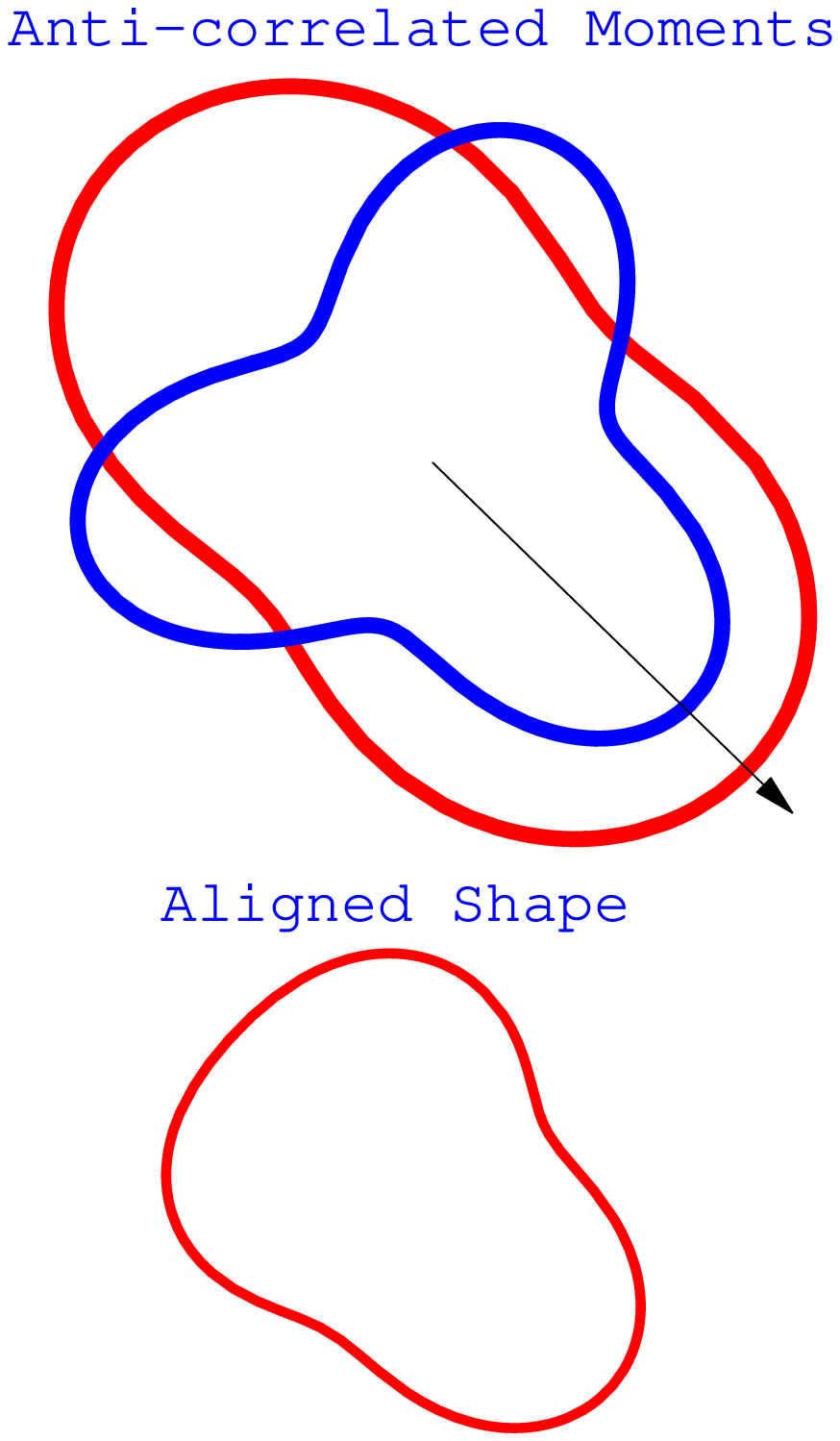}} \caption[shapes]
{Sketches showing the sum of quadrupole and sextupole moments when
moment minima are aligned (``curved'') and when moment maxima are
aligned (``aligned'').} \label{shapes}
\end{figure}
%%%%%%%%%%%%%%%%%%%%%%%%%%%%%%%%%%%%%%%%%%%%%%%%%%%
%%%%%%%%%%%%%%%% fig 12 Q-O %%%%%%%%%%%%%%%%%%%%%%%%%%
\begin{figure}[h!]
\leavevmode\epsfysize= 5.5 cm \epsfbox{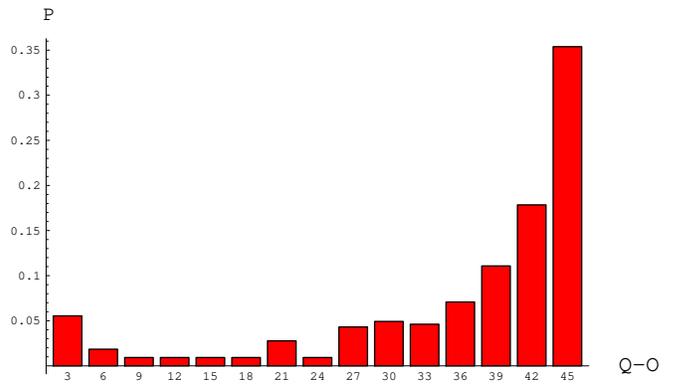}
\caption[QOCOR] {The distribution of quadrupole-octupole
orientation in the HDF  North for 324  z-selected galaxies).}
\label{QOCOR}
\end{figure}
%%%%%%%%%%%%%%%%%%%%%%%%%%%%%%%%%%%%%%

In addition to the magnitude of the sextupole moment, its
orientation with respect to the quadrupole moment is of particular
interest to us. A plot of the relative orientation of these
moments is shown in  Fig. \ref{QSCOR} for 324 z-selected galaxies
from the Hubble north field. An angle of 0$^{o}$ indicates that
one of the sextupole minima is aligned with a quadrupole minimum.
We will call such galaxies ``curved''. When the angle is 30$^{o}$
one of the sextupole maxima is aligned with a quadrupole maximum.
We shall call such galaxies ``aligned'' see Fig. \ref{shapes} .

The orientation of the octupole moment with respect to the
quadrupole is shown for the same sample in Fig. \ref{QOCOR}. The
orientation that results from an induced kick is at 0$^{o}$; the
orientation that occurs naturally, and would be present for
example in a bi-Gaussian distribution is oriented at 45$^{o}$.
Though the octupole story is of some interest, and the small bump
at 0$^{o}$ in the angular distribution is tantalizing, statistics
at this time are too small to draw meaningful conclusions and we
will not discuss the octupole further.
%%%%%%%%%%%%%%%%%%%%%%  fig 13 Q-d %%%%%%%%%%%%%%%%%%%%%%%%%%%
\begin{figure}[htbp]
\centerline{\includegraphics[width=1.7in,height=1.2in]{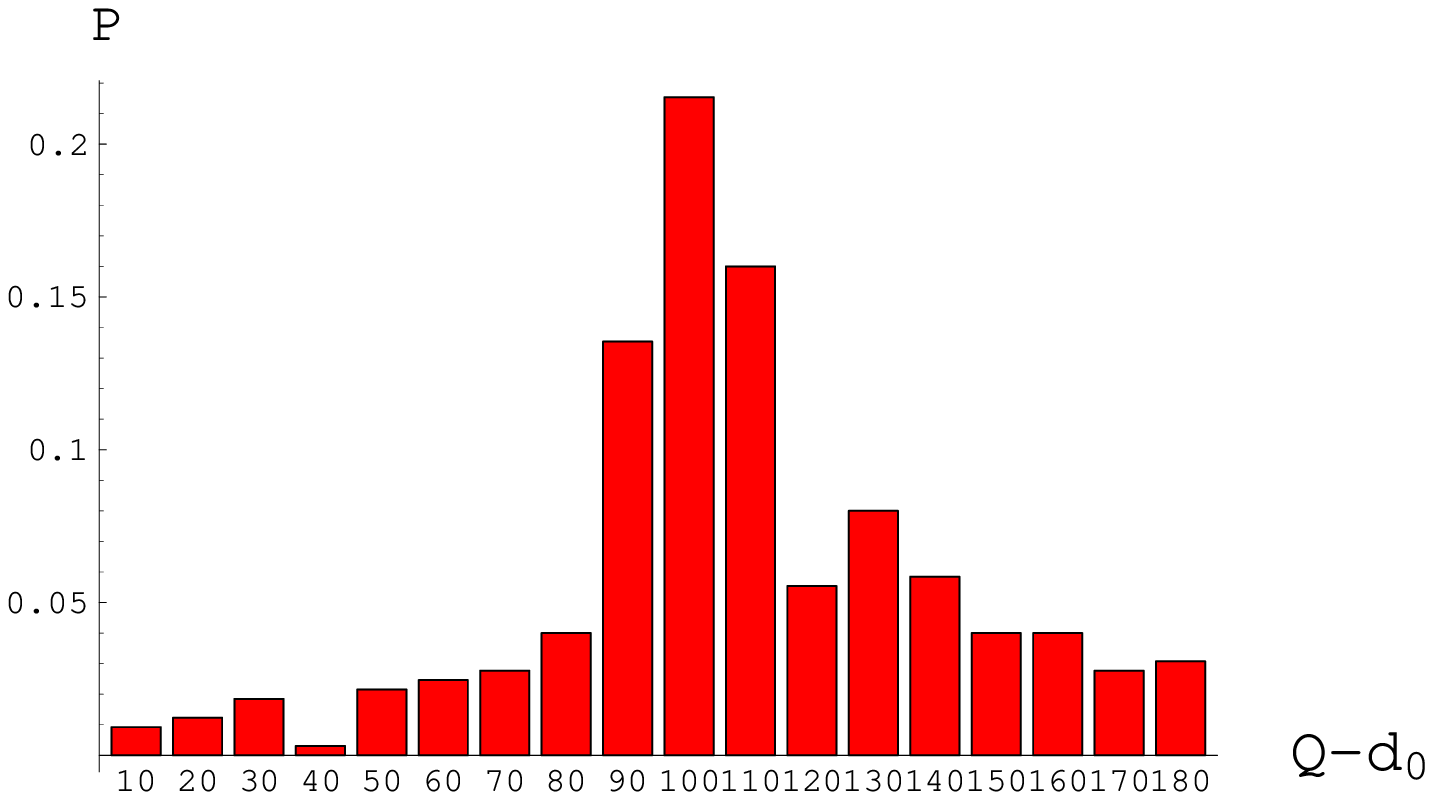}\,\,
\includegraphics[width=1.7in,height=1.2in]{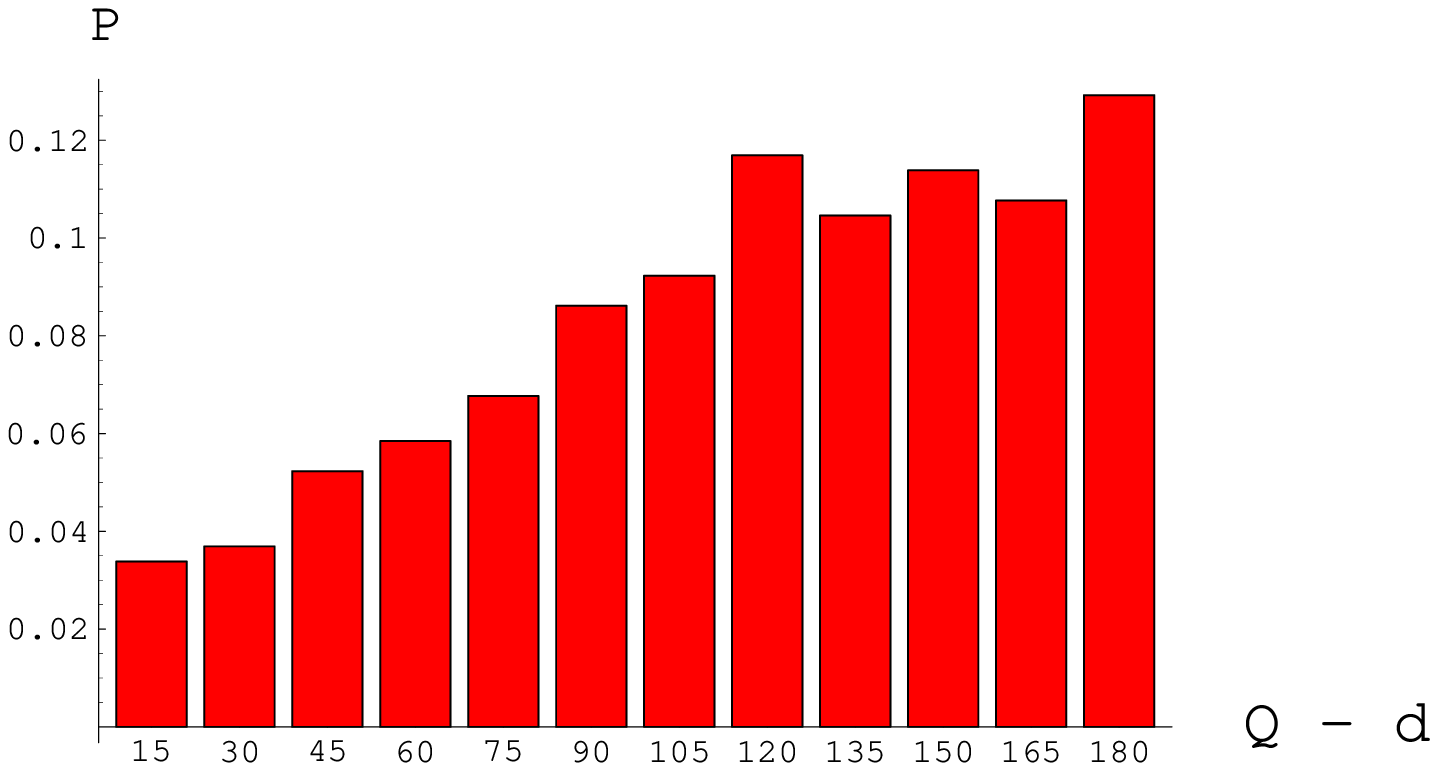}}
\caption[QdCOR] {The distribution of the orientation angle for
$d_0$ (left plot) and $d $ (right plot). $ 0^{\circ}$ is defined
by the direction to the scattering center as determined by the
quadrupole and  sextupole moment.} \label{QdCOR}
\end{figure}
%%%%%%%%%%%%%%%%%%%%%%%%%%%%%%%%%%%%%%%%%%%%%%%%%%%%%%%%%%%%%%%%%%
%%%%%%%%%%%%%%%%%%%%%% fig 13 Q-d %%%%%%%%%%%%%%%%%%%%%%%%%%%%%%%%%%
%%\begin{figure}[h!]
%%\leavevmode\epsfysize= 5.5 cm \epsfbox{hsnQd0angle.eps}
%%\caption[QdCOR] {The distribution of the orientation angle between
%%$d $ and $a. $ for HDF  North.  $ 0^{\circ}$ is defined by the
%%direction to the scattering center as determined by the sextupole
%%moment.} \label{QdCOR}
%%\end{figure}
%%%%%%%%%%%%%%%%%%%%%%%%%%%%%%%%%%%%%%%%%%%%%%%%%%%%%%%%%%%%%%%%%%%%
 Fig. \ref{QdCOR} shows the orientation of the $d $ coefficient
with respect to the quadrupole coefficient.

%%%%%%%%%%%%%%%%%%%%%%  VI %%%%%%%%%%%%%%%%%%%%%%%%%%%%%%%%%%%%%

\section{ Clumping of curved and aligned galaxies}

\begin{figure}[h!]
\leavevmode\epsfysize= 5.5 cm \epsfbox{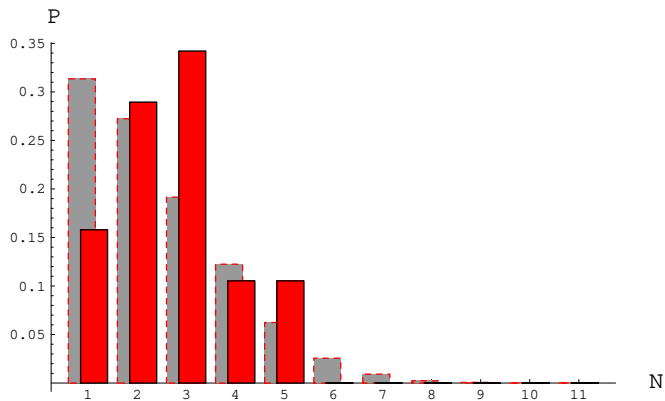}
\caption[neighbordist] {A ``neighbors'' histogram for curved
galaxies for HDF  North in the circle with  r =  360  pixels. The
background (gray) bins represent the averaged random set, the
foreground histogram corresponds to a curved set.}
\label{neighbordist}
\end{figure}

The interesting observation about ``curved'' galaxies is that they
seem to be clumped.  There are two ways to study this. The first
method is to calculate the number of curved neighbors of each
curved galaxy that lie within  a certain distance and compare this
to a distribution of the same number ($ N_c $) of randomly chosen
galaxies. We draw a circle of a fixed radius about each curved
galaxy in the field. If the circle intersects the boundary of the
field we drop the galaxy from consideration. Otherwise we count
the number of neighboring galaxies within each circle, and make a
histogram showing the number of galaxies with one, two, three,
four, and so on, neighboring galaxies inside the circle. To judge
whether the observed distribution is unusual, we have compared
this histogram with the average histogram from 300 sets of $ N_c $
randomly chosen galaxies.  Considering that the z-distribution of
galaxies might play a role in the spatial clumping, we decided to
require the randomly chosen samples to have approximately the same
z-distribution as the curved set of galaxies. An example of such a
histogram is shown in Fig. \ref{neighbordist} for the North HD
field. We call a galaxy ``curved'' if the angle between quadrupole
and sextupole minima  is between 0$^{o}$ and 8$^{o}$ (see  Fig.
\ref{QSCOR} and  Fig. \ref{shapes}).
%%%%%%%%%%%%%%%%%%%%5 fig 14 neighbors %%%%%%%%%%%%%%%%%%%%%%%
%%%%%%%%%%%%%%%%%% fig 15 different r %%%%%%%%%%%%%%%%%%%%%%%%%
\begin{figure}[h!]
\centerline{\includegraphics[width=1.65in,height=2.3in]{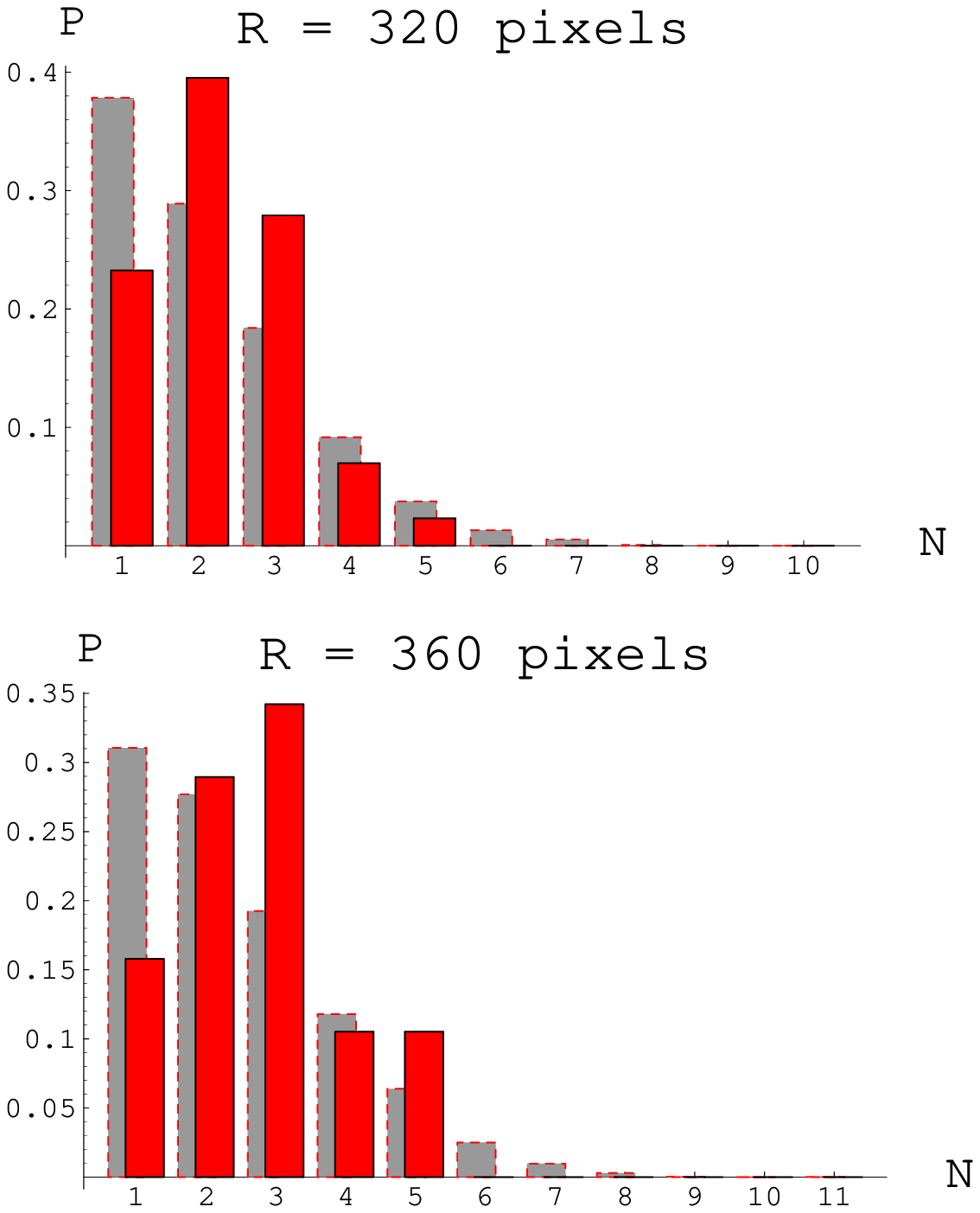}\,\,
\includegraphics[width=1.65in,height=2.3in]{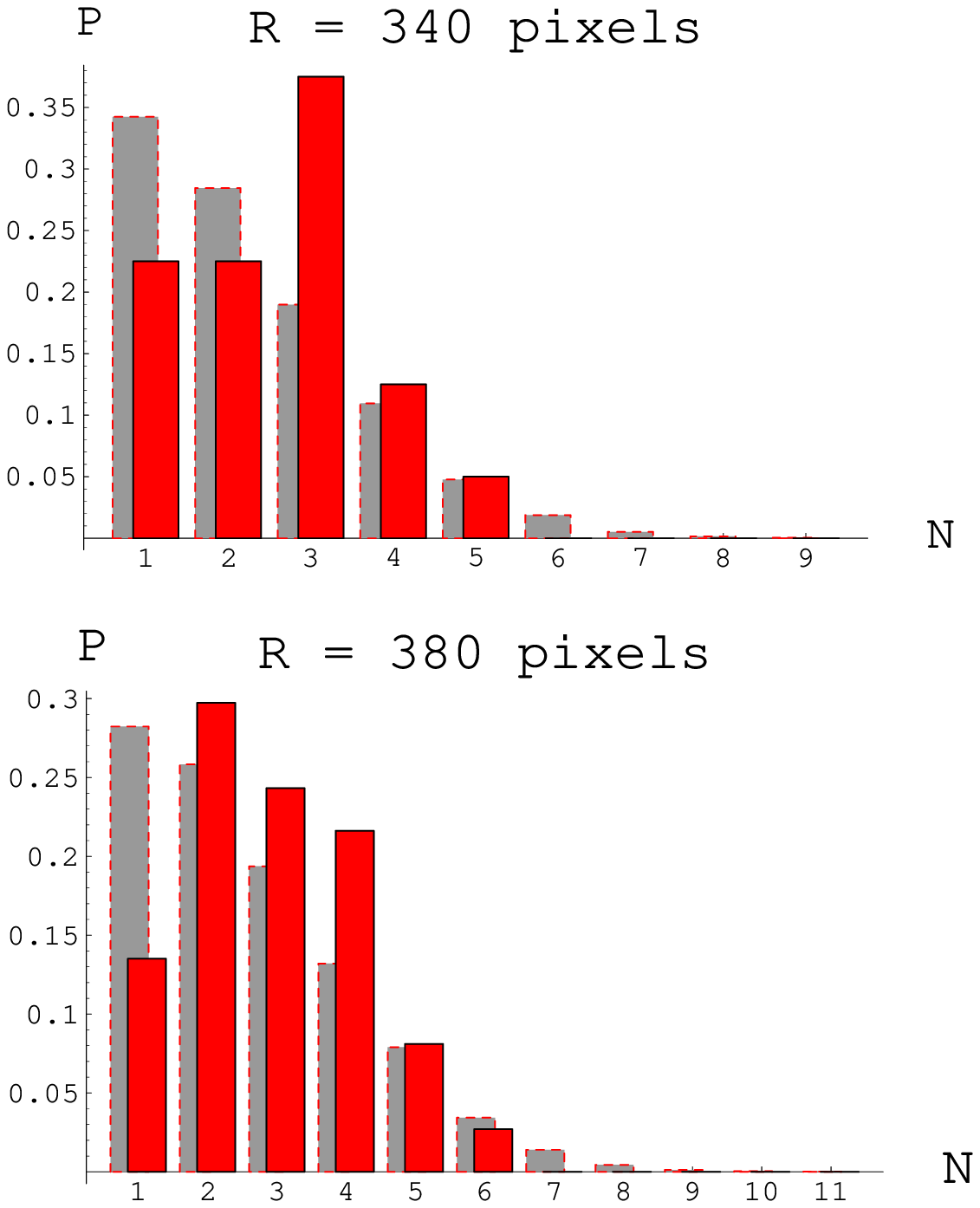}}
\caption[neighbordistrr] {A ``neighbors'' histogram for clumped
galaxies for HDF  North in the circle with  r =  320, 340, 360 and
380 pixels.} \label{neighbordistrr}
\end{figure}
%%%%%%%%%%%%%%%%%%%%%%%%%%%%%%%%%%%%%%%%%%%%%%%%%%%%%%%%%%%%%%%%%
 Fig. \ref{neighbordistrr} shows
four such histograms for several circle radii (in number of
pixels). In this radius range (320 pixels to 380 pixels) one sees
that the curved galaxy set consistently tends to have larger
numbers of neighbors within its circles.$^{ }$\footnote{ The
Hubble deep field images have a drizzled pixel size of 0.04 arc
sec. At $z$ =0.4 for current cosmological parameters (dark matter
23{\%}, baryons 4{\%}, dark energy 73{\%}) the distance scale
would be 5.6 kpc per arc sec. 360 pixels corresponds to 80 kpc. }

\begin{figure}[h!]
\leavevmode\epsfysize= 5.5 cm \epsfbox{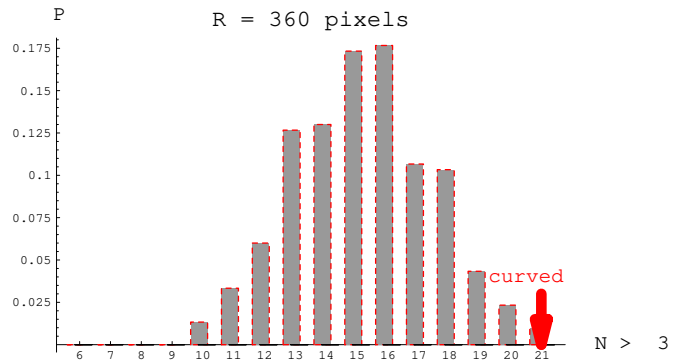}
\caption[neighbortail]{ Histogram of numbers of galaxies in 300
randomly chosen sets having 3 or more neighbors in a circle of 360
pixels. The red arrow indicates the number of galaxies with 3 or
more neighbors for the curved set.} \label{neighbortail}
\end{figure}

To determine the probability of finding such a systematic shift by
chance we compute the total number of galaxies ( in each of these
300 randomly chosen galaxy sets )  for which the circle about a
galaxy contains 3 or more neighbors. Fig. \ref{neighbortail}
%%%%%%%%%%%%%%%%%%% fig 16 tails %%%%%%%%%%%%%%%%%%%%%%%%%%
%%%%%%%%%%%%%%%%%%%%%%%%%%%%%%%%%%%%%%%%%%%%%%%%%%%%%%%%%%%%%%
displays the results of this study as a histogram: the label of
the bins indicates numbers of circles with 3 or more neighbors.
(The number corresponding to the ``curved'' set is indicated by an
arrow.) This is done for several choices of circle radius.
Typically, for the optimum radius, which is usually near 340 to
360 pixels, there will be less than 3 out of 300 sets that have as
many galaxy-circles with counts equal to or greater than the
original curved set. In other words, the probability of achieving
the curved set by chance is equal to or smaller than 1{\%}. Since
this result holds in both fields, and the fields are independent,
the probability of the observation is less than one in 10$^{4}$.

\begin{figure}[h!]
\leavevmode\epsfysize= 5.5 cm \epsfbox{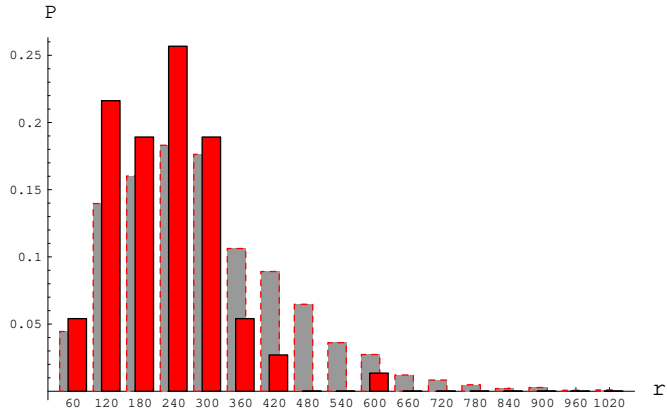}
\caption[mindis] {A distribution of the distance to the nearest
neighbor for galaxies in the north HDF.} \label{mindis}
\end{figure}

The second method to investigate the clumping is to consider a
distribution of the distances to the ``nearest neighbor'' for the
galaxies from the ``curved'' set  and compare it to the random
sets (see Fig. \ref{mindis}). The horizontal axis on Fig.
\ref{mindis} is the distance to the `nearest neighbor''.
%%%%%%%%%%%%%% fig 17 min distance %%%%%%%%%%%%%%%%%%%%%%
%%%%%%%%%%%%%%%%%%%%%%%%%%%%%%%%%%%%%%%%%%%%%%%%%%%%%%%%%%%%%%
The Kholmogorov-Smirnov test gives  $ 0.5\% $  probability that
these two distributions are the same. But the situation is
actually less probable than that, for two further reasons.
%%%%%%%%%%%%%%%%% fig 18 aligned min dist %%%%%%%%%%%%%%%%%%%%%%%
\begin{figure}[h!]
\leavevmode\epsfysize= 7.5 cm \epsfbox{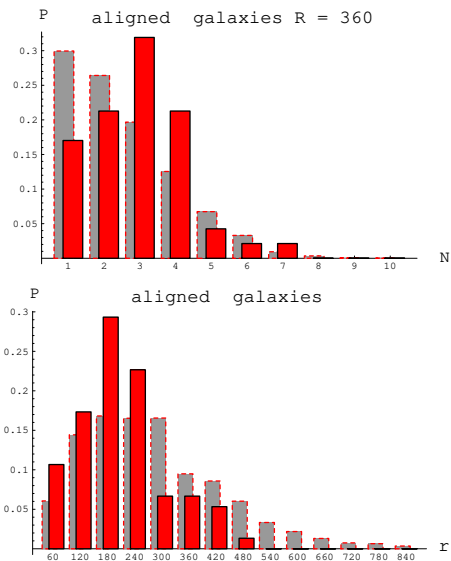}
\caption[mindistac] {A ``neighbors'' histogram for aligned
galaxies in a circle with  r =  360  pixels and the average
histogram for randomly chosen sets in the north HDF.}
\label{mindistac}
\end{figure}
%%%%%%%%%%%%%%%%%%%%%%%%%%%%%%%%%%%%%%%%%%%%%%%%%%%%%%%%%%%%

Given any distribution of background galaxy orientations, the
lensing hypothesis would suggest that if the regions that are
populated by dark matter clumps are giving rise to the clumping we
observe, then equally well, the voids should give rise to a
population of galaxies that are ``straighter'', more aligned than
normal. In other words, we would predict clumping of the aligned
galaxies (Fig. \ref{shapes} ). We have carried out the same
procedure as for the ``curved'' galaxies for ``aligned'' ones and
found that the ``aligned'' galaxies are even more clumped (Fig.
\ref{mindistac}). The probability is less than  1 {\%}.

%%%%%%%%%%%%%%%%%%%%% Z distribution %%%%%%%%%%%%%%%%%%
\begin{figure}[h!]
\leavevmode\epsfysize= 4cm \epsfbox{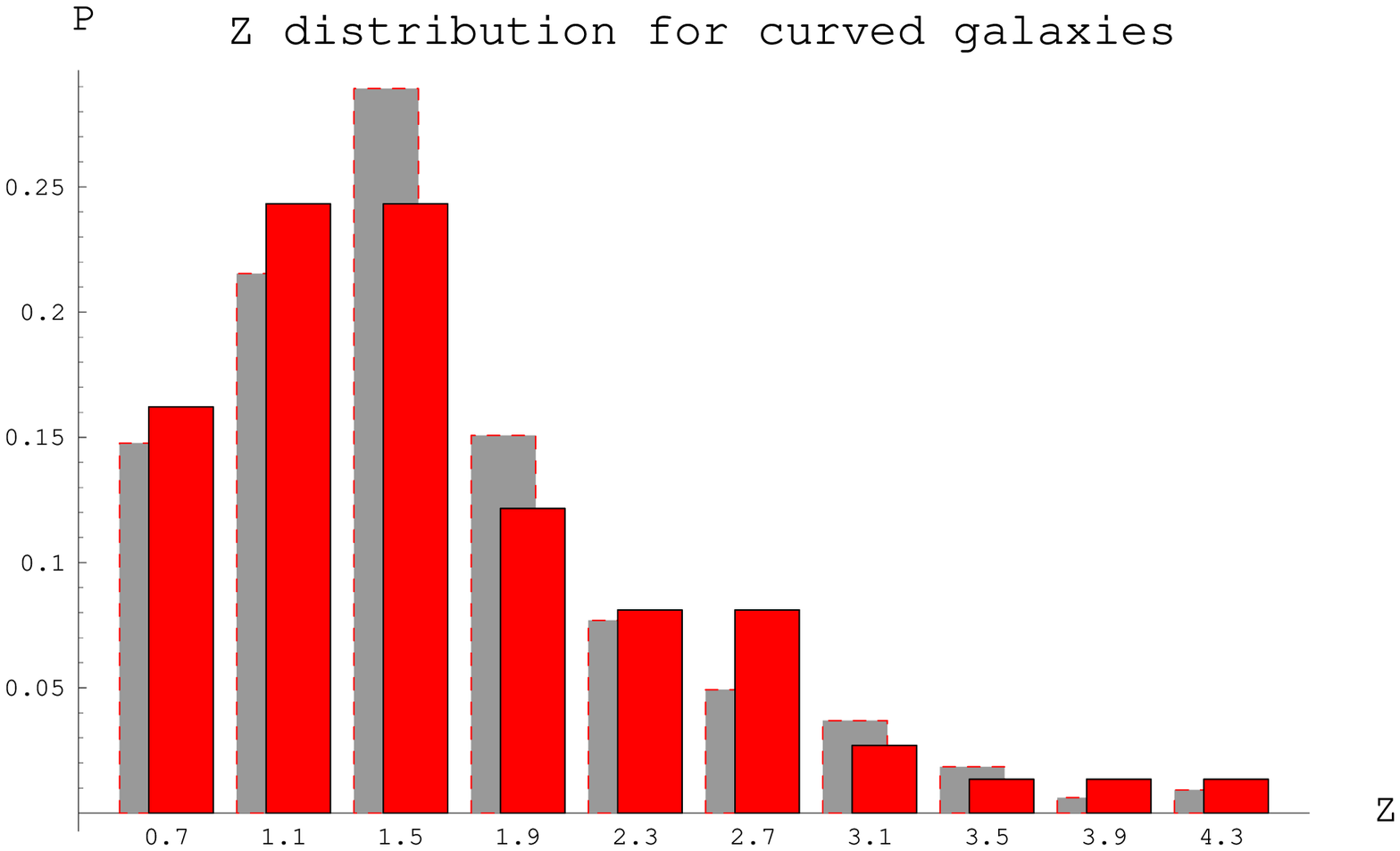} \caption[Zcc]{The
red (grey) bars indicate the z-distribution of ``curved'' (all)
galaxies, respectively .} \label{Zcc} \leavevmode\epsfysize= 4cm
\epsfbox{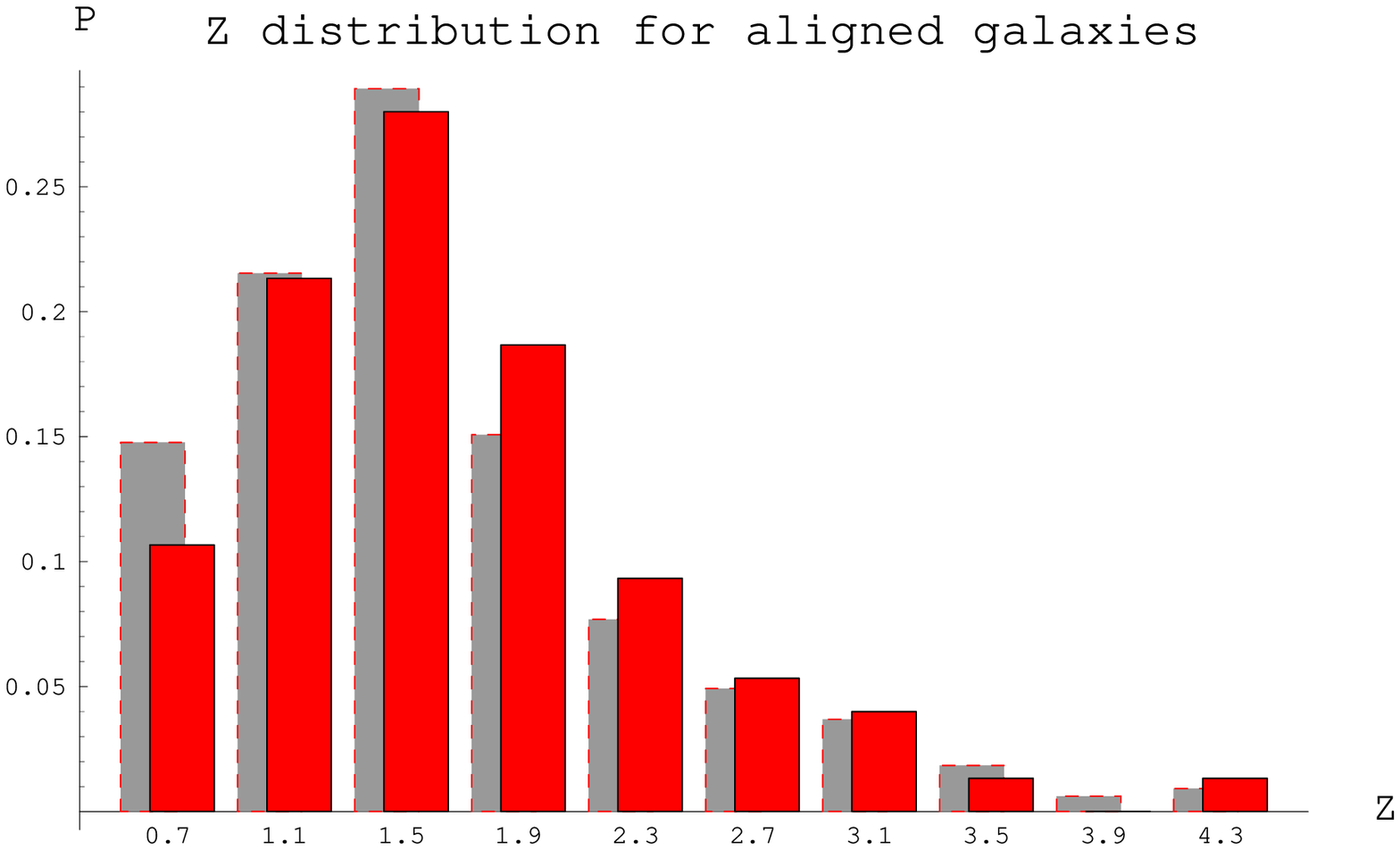} \caption[Zac]{The red (grey) bars indicate
the z-distribution of ``aligned'' (all) galaxies,  respectively..}
\label{Zac}
\end{figure}
%%%%%%%%%%%%%%%%%%%%%%%%%%%%%%%%%%%%
That aligned galaxies are also clumped is not a trivial
consequence of the fact that curved galaxies are clumped. Indeed
galaxies midway between curved and aligned are not clumped. And
though the aligned set is not completely independent of the curved
set (it represents $\raise.5ex\hbox{$\scriptstyle 1$}\kern-.1em/
\kern-.15em\lower.25ex\hbox{$\scriptstyle 4$} $ of the complement
of the curved galaxies), we maintain that it is independent enough
to assert that the probability of finding both clumped by chance
is the product of the probability of each. With less than a 1{\%}
probability of curved galaxies being clumped and a 1{\%}
probability for aligned galaxies being clumped in both the north
and south fields, the chance probability of all four events is
less than 1 in a million.\footnote{ We are not yet claiming the
clumping is due to lensing, only that there is clumping.}

%%%%%%%%%%%%%%%%%%%%%% NHDF  - galaxies %%%%%%%%%%%%%%%%%%%%%%%%%%%
\begin{figure*}[t]
\centerline{\includegraphics[width=3.5in,height=3.5in]{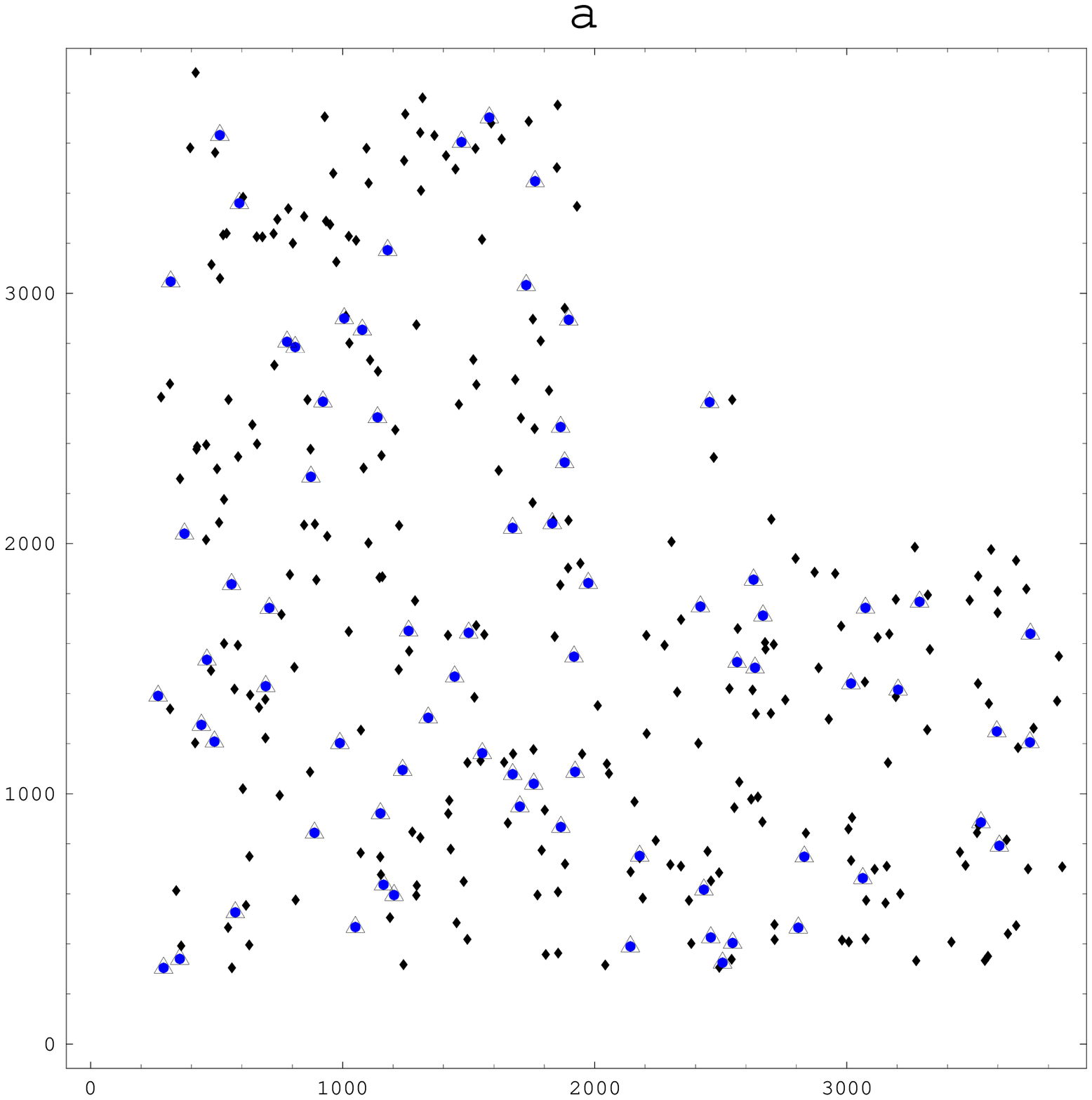}\,\,
\includegraphics[width=3.5in,height=3.5in]{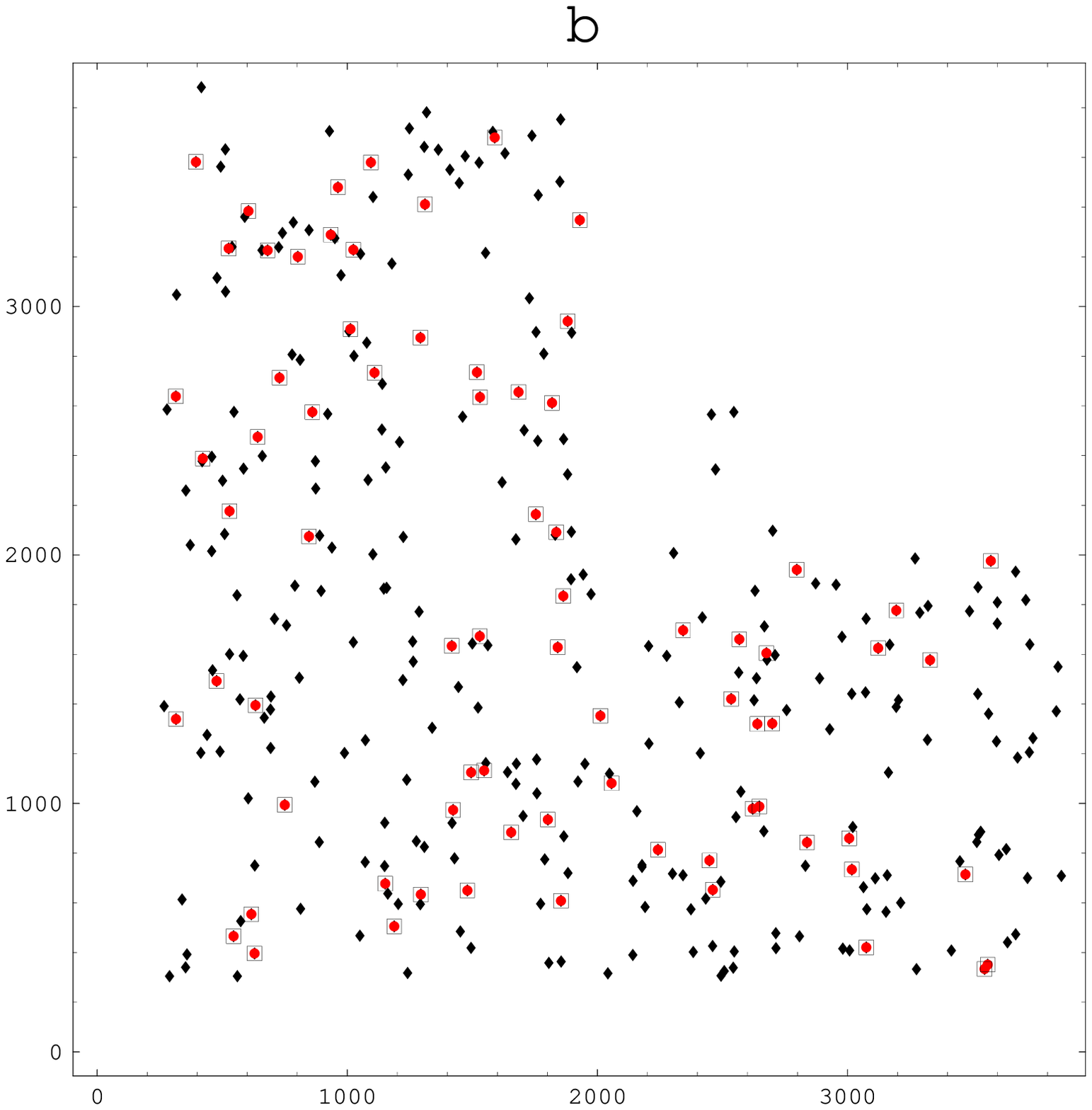}} \caption[fieldaccc]
{The HDF (North):(a) -- triangles (blue dots) are curved galaxies,
(b)-- boxes (red dots) are ``aligned'' galaxies and black dots are
all other background galaxies.} \label{fieldaccc}
\end{figure*}
%%%%%%%%%%%%%%%%%%%%%%%%%%%%%%%%%%%%%%%%%%%%%%%%%%%%%%%%%%%%%%%%%%

Fig. \ref{Zcc} shows and compares the z-distribution of curved
galaxies with the z-distribution of all background galaxies. Fig.
\ref{Zac} compares the z-distribution of aligned galaxies with the
z-distribution of all  background galaxies. Even though the
z-distributions for  both curved and aligned sets
 are not significantly different from the  average distribution we
have checked the influence of z-dependence on the observed
clumping. We ran the procedures described above for two cases:
first, for the random sample restricted to mimic the
z-distribution of a curved (aligned) set and, second, for
z-``blind'' random sets. The results for the probability of
clumping for both cases were similar. $^{ }$\footnote{We have
found that when the randomly chosen set is required to match the
z-distribution of the curved set (in 8 or 10 z-bins) then the
statistics tests give 1{\%}  probability of getting clumping
 by chance compared to  3{\%} probability for the z-independent choice.}
This result was confirmed for both fields. We have also carried
out an identical process, with similar results, for galaxies
selected by magnitude rather than z-value. As a result of these
considerations and observing the persistence of clumping under a
wide variety of circumstances, we are confident that the clumping
we observe is not occuring by chance. In the next section we
discuss several possible alternative sources of clumping. In Fig.
\ref{fieldaccc}(a) we show the spatial location of the ``curved''
galaxies of the north field among all remaining background
galaxies.  In Fig. \ref{fieldaccc}(b) we show the spatial
distribution of ``aligned'' galaxies of this field with all other
background galaxies.

%%%%%%%%%%%%%%%%%%%%%%%%%%%%%  VII %%%%%%%%%%%%%%%%%%%%%%%%%%%%%%%%%%%%%%%

\section{ Alternative explanations of clumping}

Though we are comfortable in asserting that the clumping results
cannot occur by chance they may be due to reasons other than
lensing:

\begin{enumerate}
\item Instrumental
    \begin{enumerate}
    \item The point-spread function varies over the field
    \item Pixel derived effects
    \end{enumerate}
\item Derived from the clumping of the background galaxies
themselves
    \begin{enumerate}
    \item Time evolution of background galaxy groups
    \item Some other group property
    \end{enumerate}
\item Computational
    \begin{enumerate}
    \item The coupling of quadrupole and sextupole through $M_{21}^T $.
    \item Galaxy selection and analysis methods
    \item Image composition (drizzling, overlay, etc.)
    \end{enumerate}
 \end{enumerate}

We now address each of these concerns.

{\bf 1(a)}. The point-spread function could presumably have a
sextupole and quadrupole moment and these could be aligned with
one another in some smoothly varying way across the image. However
we feel that this is ruled out by the facts that:
\begin{table*}[htbp]
\caption{  Alternative explanations for clumping of curved
galaxies. Notations: +  $\Rightarrow$ supports hypothesis; OK
$\Rightarrow$ neutral; \\ NO $\Rightarrow$ contradicts hypothesis;
- $\Rightarrow$ contradicts but more data required; NA
$\Rightarrow$ not applicable. }
\begin{tabular}{|p{140pt}|l|p{69pt}|p{72pt}|p{68pt}|p{73pt}|}
\hline Observations & Lensing& 1(a) PSF & 2(b) Group property &
3(a) $ M_{21}^T $  \par &
3(c) Image composition \\
\hline 1. Clumping of curved & + & OK & OK&
 NO &
 - Only along  \par boundaries \\
\hline 2. Clumping of aligned& + & OK& OK & NO &
 - Only inside \par boundaries \\
\hline 3. No clumping of \par  mid-range galaxies & + & -
  Could  clump&
  OK &  OK&
OK \\
\hline 4. Mixed z content of curved clumps & +& OK & -  & OK&
OK \\
\hline 5. Curved and clumped concentrated at small a,b.&  + & OK &
OK & NO \par  Prefers large $a$ &
OK \\
\hline 6. Aligned and clumped not concentrated in a,b.& + & OK&
OK& OK &
OK \\
\hline 7. Direction to scattering center varies & + & -  & OK& OK&
NO  \par Pattern expected \\
\hline 8. Curved galaxies next to not-curved galaxies & + & NO &
OK & OK &
NO \par Pattern  expected \\
\hline 9. Stars are round & OK & - \par Not enough stars & OK & OK
&
-  \par Not enough stars \\
\hline 10. Cluster CL0024& + & \textit{NA}& \textit{NA}&
\textit{NA} &
\textit{NA} \\
\hline 11. Deduced mass magnitudes have reasonable values& + &
\textit{NA}& \textit{NA}& \textit{NA}&
NA \\
\hline
\end{tabular}
\label{tab1}
\end{table*}

\begin{enumerate}
\item the 6 stars in the Hubble north field show no sign of having
a quadrupole or sextupole strength of the required magnitude;
\item mid-range galaxies (not aligned and not curved) could just
as well have been clumped under this scenario and are not; and
\item the deduced directions to the scattering center are erratic.
\end{enumerate}

The logic behind 2) is that if the point-spread function is
causing this effect, its sextupole moment and quadrupole moment
are continuously varying across the image so as to induce the
observed distortion in the images. There could be isolated areas
of the sky were this effect is mid-range (not necessarily aligned
or curved).

{\bf 1(b)}. We have carried out studies to see if unexpected
moments are generated by dividing an image into pixels. For
example, we took a known bi-Gaussian distribution and, varying the
centroid, projected it onto a pixel grid. The falsely induced
sextupole moment had strengths less than 10$^{-5}$. In general,
there is no reason we have discerned for which pixelation effects
lead to spatial correlations, since pixels (modulo pixel defects)
are uniform across the field.

{\bf 2(a)}. Since the galaxies at a slice in $z$ are known to be
clumped, then if there is some age-dependent change in galaxy
shapes, a shape selection criteria could be seeing an age-biased
sample which could then be clumped. Since our ``curved'' galaxy
set has the same $z$-distribution as all galaxies, the premise
would appear to be false, ie galaxy curvature as we are
quantifying it does not appear to evolve with time.

{\bf 2(b)}. The premise is that galaxy groups possess some
property (other than age) that effects the shape of galaxies.
Perhaps there was an explosive event or set of events that
extended across the group (80 kpc radius) that effected the birth
process of the background galaxies. If true, it would be an
unexpected and interesting result in itself. We don't think we can
rule this out yet, but with more statistics an argument to
discount this scenario could be based on the fact that the curved
and clumped galaxies in any given clump have a variety of
z-values.

{\bf 3(a)}. If $M_{21}^T $ is non-zero, a $\Delta M_{20}$ can lead
to a correlated quadrupole and   sextupole moment.  A clumping of
``curved''  galaxies could result from a spatial dependence of the
orientation of  $M_{21}^T $ with respect to $\Delta M_{20}.$ A
non-lensing origin of such a spatial dependence is an alternative
source of clumping, but would require its own explanation,
presumably one of the other alternatives among the items in our
list.

{\bf 3(b)}. Galaxy selection would appear to be blind to position
in space and incapable of leading to the spatial clumping of
curved or aligned galaxies

{\bf 3(c)}. This represents our most serious concern. We suspect
that we are processing this data in a way which was not
anticipated by the creators of the image construction process, and
we are concerned, for example, that on the boundaries of overlays,
distortions could be present that would give the images a false
shape. There are not enough stars in the field to rule this out
using star images. Our best response to this concern is to note
that the regions where curved galaxies clump do not appear to have
any particular identifiable pattern, i.e. they do not appear to
coincide with sub-field boundaries.

%%%%%%%%%%%%%%%%%%%%%%%%%%%%%  VIII %%%%%%%%%%%%%%%%%%%%%%%%%%%%%%%%%%%%%%%
\begin{figure*}[t]
\centerline{\includegraphics[width=3.5in,height=2in]{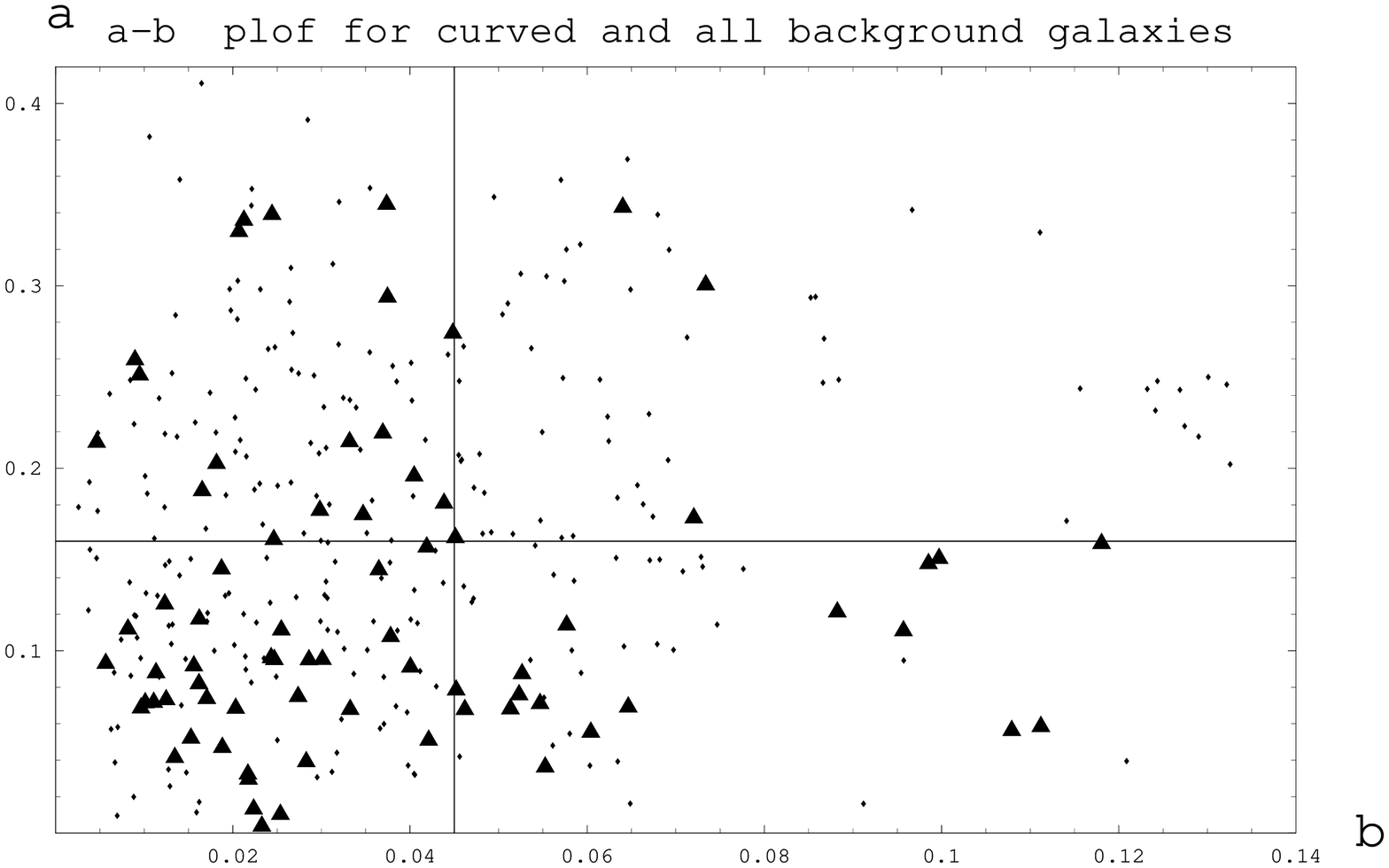}\,\,
\includegraphics[width=3.5in,height=2in]{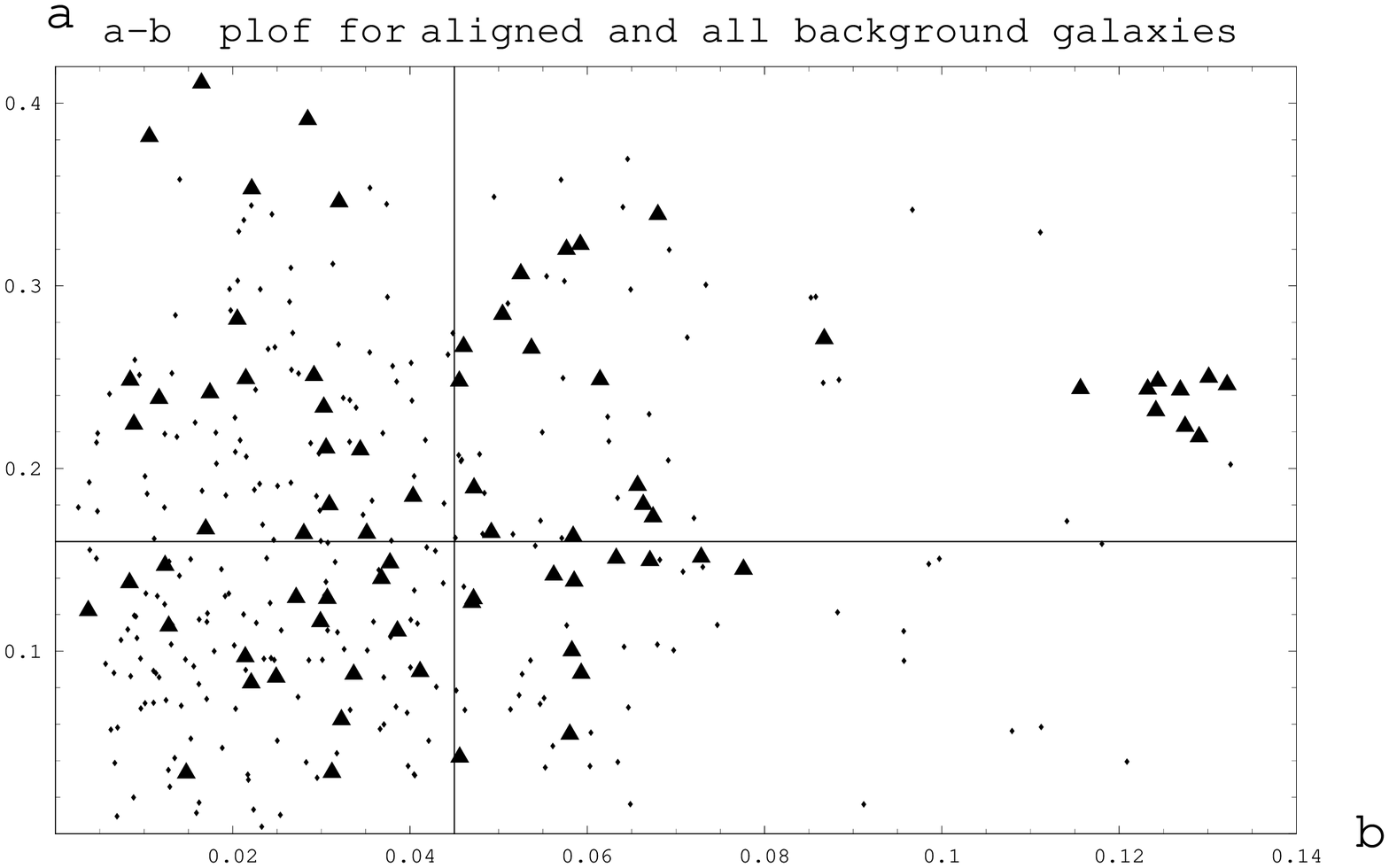}} \caption[aabb]
{The HDF (North): triangles correspond to ``curved'' (left plot)
and  ''aligned'' (right plot) galaxies and black dots on both
plots are all other background galaxies.} \label{aabb}
\end{figure*}

\section{The ${\bf \large a-b }$  moment-magnitude plane}

Fig. \ref{aabb} (left) is what we call an $\left( {\hat {a},\hat
{b}} \right)$ plot for the sample of all galaxies (dots) and the
set of curved galaxies (triangles). Each point on the plot
corresponds to a single galaxy. The horizontal axis is the
magnitude of the deduced sextupole coefficient $b$, and the
vertical axis is the magnitude of the deduced quadrupole
coefficient $a$. Note that galaxies having both large $a$ and
large $b$ which occur in the set of all galaxies are noticeably
absent in the set of curved galaxies. In the $\left( {\hat
{a},\hat {b}} \right)$ plot for all galaxies 40{\%} have both $a$
and $b$-values below the mean, whereas in the $\left( {\hat
{a},\hat {b}} \right)$ plot for curved and clumped galaxies,
70{\%} have both $a$ and $b$ values below that same mean.

This trend was predicted for lensing, as a consequence of
considering the vector addition of the original moment and the
induced moment. When the original moment is larger than the
induced moment, the resulting vector tends to be aligned in the
direction of the original moment, and since this angle is then
divided by 2 in the case of the quadrupole moment, and by 3 in the
case of the sextupole moment, the alignment with the original
moment is much better than would be normally expected. On the
other hand when the induced moment is larger than the original
moment the argument works the same way to deduce that the
alignment with the induced moment is much better than expected.

In other words, the induced curving is expected to be seen if the
background moments are as small as, or smaller than the induced
moment. So the population on the $a-b$ plot for curved and clumped
galaxies is a strong indication of the strength of the induced
moments. Of course, any other ``add-on'' effect will have the
characteristic that it will be more noticeable for small normal or
original moments, but the vector addition law with the
\textit{2$\theta $} and 3\textit{$\theta $} dependence has a
remarkably sharp behavior.

``Aligned''  galaxies, on the other hand, are presumed to
represent galaxies that have not been altered by lensing into a
mid-range or curved shape.  Thus their distribution on the a-b
plane should be very similar to all galaxies.  This is indeed the
case, as can be seen in Fig. \ref{aabb} (right plot).

%%%%%%%%%%%%%%%%%%%%%%%%%%%%%  IX %%%%%%%%%%%%%%%%%%%%%%%%%%%%%%%%%%%%%%%

\section{Estimate of group mass}

The results of this and the following section are more
speculative. We assume that the observed clumping is indeed due to
lensing and attempt to deduce the properties of the lensing mass
distribution. This is premature because we have not done the
modeling to determine systematics and because the sample of
clumped and curved galaxies is small (total of 110, both fields).
On the other hand, we feel it is possible and appropriate to make
order of magnitude estimates.

An estimate for the mass $M_X $ of the group could be written \BE
M_X  = \rho \,A_X =\frac{M}{d_X^2 }A_X =\frac{M}{r_0^2 }\left(
{\frac{r_0 }{d_X}} \right)^2 A_X \,, \EE where $M$ is the mass of
the constituent clumps, $A_X $ the area of the group, and $d_X $
the typical separation distance of the clumps within the group.
$r_0^2 $ is chosen so that insertion of $M/r_0^2 $ into the
formula for $a$ produces a typical induced moment size for events
that change the population of curved galaxies. $P_a \equiv \pi
r_0^2 /d_X^2 $ can be interpreted as the probability that any
particular light path receive a moment change of the required
magnitude. This can be found by obtaining an estimate for the
probability that the light path of any particular galaxy passing
through the group receives noticeable induced moments.

\begin{figure*}[t]
 \centering\leavevmode\epsfysize= 14cm \epsfbox{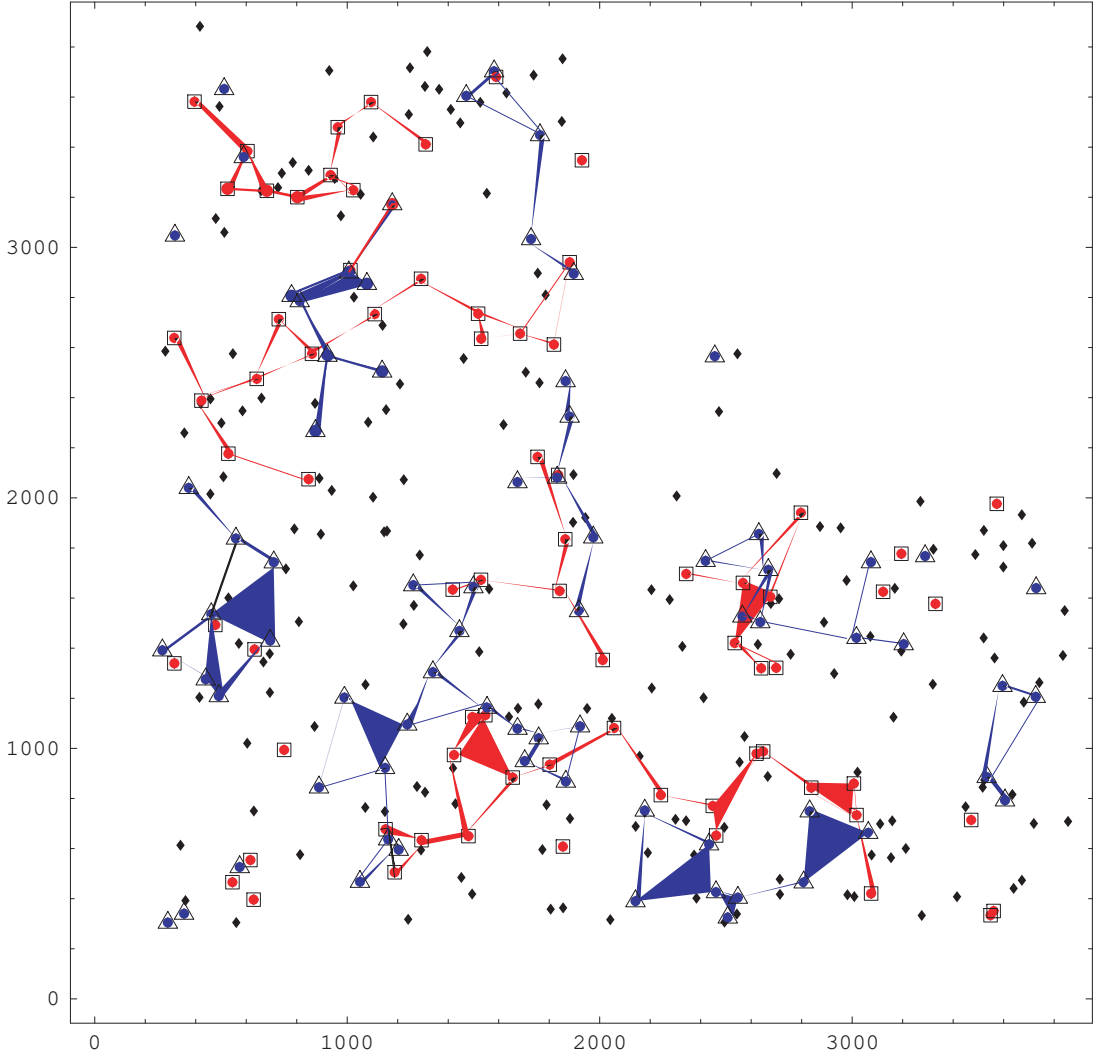}
 \caption[field] {The HDF  North  red boxes are ``aligned''
 galaxies, blue  triangles are curved galaxies and black dots are all
other background galaxies.} \label{field}
\end{figure*}

To obtain this probability we have attempted to identify the
clumped parts of the field (which is justified since we know it is
indeed clumped) and determine the fraction of the galaxy light
streams which become curved when penetrating these parts.  We have
done this using  a friends-of-friends algorithm, requiring that i)
there are at least 4 curved members in a group, and ii) that to be
included in a group a curved galaxy must lie within a certain
distance (about 350 pixels) of other members of the group. Fig.
\ref{field} is a combination of fig. \ref{fieldaccc}(a)  and fig.
\ref{fieldaccc}(b) with the ''curved'' and aligned galaxies
identified as belonging to clumps connected by straight lines.
%%% \epsfbox{fieldNd8ccacLine4V2.ps}
We note that we have groups with area ranging from 6 10$^{5}$
pixels to 1.5 10$^{5 }$ pixels. Within the groups about 40{\%} of
the galaxies are curved. Outside of the groups less than  10{\%}
of the galaxies are curved. So the probability of being
transformed to curved is estimated at 30{\%}.

The transformation to curved depends on the suitability of the
background galaxy (for example its moments are small enough to
easily alter) times the probability that the kick is large enough.
We assume that these are roughly equal, and hence get a
probability of $P_a =0.6$ that the kick is large enough.

This yields the estimate \BA M_X G & = & \frac{a_{typ} }{4\pi
}\frac{D_{TS} }{D_{TL} D_{LS} }\;P_a A_X  \\
\nonumber \\
  & \approx & (0.5 \mbox{
to 2) }a_{typ} [{\rm pc}]. \nonumber \EA
 $a_{typ} $ has to be
large enough to change moment alignments, so it must be comparable
to but probably less than $\sigma _a $. Also we must account for
the systematic weakening of moments by the point-spread function,
which (see below) we would estimate at $a\approx 2\hat {a}$. For
$a_{typ} \approx 0.2$, $M_X G\approx (0.1\mbox{ to
0.4)}\;pc\Leftrightarrow M_X =(2 \mbox{ to 8
10}^{\mbox{12}})\;M_\odot $.

Moments $a $ cannot be larger than $\sigma_a$ because the
distribution of moment strengths for curved galaxies is actually
smaller than that of all galaxies. The realignment of moments can
be thought of as a change of direction of the quadrupole moment or
a change of direction of the sextupole moment, or a combination of
both.

A distinct possibility is that there is no induced sextupole; only
the quadrupole moment direction is changed leaving a sextupole
behind resulting in a ``curved'' galaxy. One might think that
higher order moments get moved along with the quadrupole because
the map is linear. For example a bi-Gaussian, which has its
octupole moment aligned with the ellipticity, is mapped into a
bi-Gaussian which will also have its octupole moment aligned with
its ellipticity. However, for sextupole moments this question is
answered quantitatively by equation (\ref{eq11}) which includes a
term showing that with $b=0,$  the observed sextupole moment
change will depend on the magnitude of $M_{21}^T $.

%%%%%%%%%%%%%%%%%%%%%%%%%%%%%  X %%%%%%%%%%%%%%%%%%%%%%%%%%%%%%%%%%%%%%%

\section{Estimate of clump mass}

The typical groups in the field could presumably be composed of 5
clumps of mass 10$^{12}M_\odot $, 50 clumps of mass
10$^{11}M_\odot $, 500 clumps of mass 10$^{10}M_\odot $, 5000
clumps of mass 10$^{11}M_\odot $, or some fractional combination
of these. A single mass at 5 10$^{12}$ solar masses is ruled out
by the erratic directions to the scattering center. We are left
with this degeneracy because the expression for the induced
quadrupole moment of equation (\ref{eq5}) determines only the
ratio  $M/r_0^2 $. Estimating the clump mass requires additional
information.

Additional information can come from an analysis of the relative
role played by the quadrupole and sextupole moments, according to
the following considerations. It follows from equation (\ref{eq5})
that the strength of the induced sextupole moment occurring along
with the induced quadrupole moment $a_{typ} $ would be $b_{typ}
=a_{typ} \;r_G /r_0 $. Assuming the lensing plane is at $z=0.4$,
$r_G \approx 0.5\;kpc$ and since  $\sigma _b /\sigma _a \approx
1/5$, and as we will argue below that $a/b\approx \hat {a}/\hat
{b}$, then both sextupole and quadrupole moments will be equally
observable for an $r_0 \approx 2.5\;kpc.$ Since we have estimated
above that $P_a =\pi r_0^2 /d_X^2 \approx 0.6$, for this $r_{0 }$
we have  $d\approx 6\;kpc$  and $ N=A_X /d_X^2 \approx 500$ for
the typical groups.  Thus for $b_{typ} =1/5\;a_{typ} $  the
estimate for the clumping mass would be $M\approx 10^{10}\;M_\odot
$. Following this line of reasoning we have constructed the
following table.

\begin{table}[htbp]
\caption{ Possible lensing parameter range. }
\begin{tabular}{|p{107pt}|l|l|l|l|}
\hline Lensing mass /$M_\odot $& 10$^{9}$& 10$^{10}$& 10$^{11}$&
10$^{12}$ \\
\hline Number of clumps in largest groups & 5000 & 500 & 50 &
5 \\
\hline Typical impact parameter $r_{0 }$\textit{(kpc)}&
0.8&
2.5&
8&
25 \\
\hline $b/a$ ratio&
0.6& 0.2& 0.06&
0.02 \\
\hline $d_{X}$\textit{ (kpc)}& 2& 6& 20&
60 \\
\hline
\end{tabular}
\label{tab2}
\end{table}

From a lack of a population of localized large $b$ values, masses
below  10$^{9}$  are ruled out. Under more careful inspection and
better statistics the range from a few 10$^{9}$ through 10$^{11}$
should have observable consequences for $b$. The presence of
lensing masses greater than 10$^{11}$ would be hard to numerically
constrain from looking at sextupole distributions, except to know
the mass is larger than the observable cut-off. However, there
should be observable consequences for the quadrupole distribution,
since there would be a population of larger induced moments.

In other words, we believe that a careful study of the
distributions of $a$ and $b$ together with modeling should be able
to provide sufficient statistical information to deduce the
distribution of lensing masses.

We end this section with a discussion of systematic and random error.
Important error and noise sources in any individual measurement include:

\begin{enumerate}
\item the systematic effects of the point-spread function and thresholding,
\item random background galaxy moments,
\item extraction of moments from the image,
\item pixelation of images, and
\item photon counting noise.
\end{enumerate}

(1)  The largest effect arises from thresholding and the PSF. For
our galaxy cores, we estimate $M_{11}^P \approx M_{11}^T $, whence
$M_{11}^T \approx \frac{1}{2}\hat {M}_{11}^T $, implying $a\approx
2\,\hat {a}$. For Gaussian shapes, $M_{22}= 4 M_{11}^2$, hence an
estimate for the denominator of the expression for $\hat {b}$ is
approximately $10\;\sigma ^4\sim 2.5\;M_{22}^T $ and the numerator
has an $\sqrt {\hat {M}_{11}^T } \sim \sqrt {2M_{11}^T } $,
yielding the estimate $b\approx 2\,\hat {b}$. These are large
adjustments that need careful attention, but the gross indication
is that the $a/b$ ratio will suffer a smaller adjustment than
either numerator or denominator and led to the use above of an
estimate $a/b\approx \hat {a}/\hat {b}$.

(2) The next largest uncertainty in the $\hat {a}/\hat {b}$ ratio
comes from the existence of non-zero background moments. It is a
happy circumstance that there is clumping of ``aligned galaxies''
as well as ``curved'' galaxies, because the clumps of aligned
galaxies are presumed to indicate a lack of lensing within their
domains. Hence the distribution of galaxies behind voids will give
an important base to which the galaxies behind lensing clumps can
be compared. With larger statistics, one could expect to extract
interesting details by studying these differences. Also it will be
invaluable to have a precise knowledge of the point-spread
function, including all its moments, throughout the field.

(3) We feel that our method to determine moments, validated by its
ability to reveal important correlations, can be substantially
improved. This is intimately related to item (1).

(4) In  pixelation studies, we were surprised to find that the
change of the centroid distribution was unable to produce apparent
sextupole moments above the 10$^{-5}$ level. Square pixels do give
rise to spurious octupole moments, and some care is required in
that case.

(5) Our thresholds are typically set at approximately 175 photons
per pixel. The core peaks are the order of 600 photons counts per
pixel. We have not studied the effects of photon counting noise,
however we would not expect its obvious random nature to affect
our conclusions, and it should be substantially smaller than item
(2).

%%%%%%%%%%%%%%%%%%%%%%%%%%%%%  XI %%%%%%%%%%%%%%%%%%%%%%%%%%%%%%%%%%%%%%%

\section{Determination of clump radii}
\label{sec:determination}

The probability of penetrating a clump would go like
\begin{equation}
\label{eq19} \mbox{Probability}=\frac{\pi r_L^2 }{d_X^2}=\frac{\pi
r_L^2 }{M}\frac{M}{r_0^2 }\frac{r_0^2 }{d_X^2}=\frac{\pi r_L^2
}{M}\frac{M_X }{\pi r_X^2 },
\end{equation}
where $r_L $ is the lensing mass (clump mass) radius. The last ratio in this
equation is fixed, so the probability depends only on the first ratio (the
inverse of the projected density within a clump) which could be slowly
varying. While one expects the 3-dimension density of small clumps to be
larger than large clumps, the projected densities can be similar. So the
probability of penetrating a clump could depend very weakly on the
composition of the group.

\begin{figure}[h!]
\leavevmode\epsfysize= 5.5cm \epsfbox{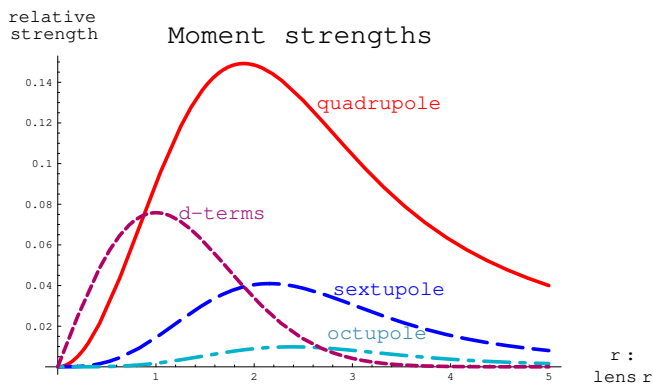}
\caption[strength] { Strength of the quadrupole, sextupole,
octupole, and cardioid-like kicks within a Gaussian clump. The
horizontal axis is the radius divided by the RMS lens radius. }
\label{strength}
\end{figure}
Penetration within a clump mass distribution will affect the
strength of the induced moments and should leave a signature on
the sextupole and quadrupole moment distributions. This effect is
larger than one might initially expect because the sextupole
moment within a clump is strongly suppressed and only reaches its
asymptotic form at $r=3r_L .$ Fig.\ref{strength} illustrates the
situation fully for a Gaussian mass distribution. If $r_L \ge r_G
$ ($r_{G}$ is the background galaxy light-stream footprint radius)
then the interior can be well probed, and the quadrupole to
sextupole ratio will remain less than
\[
\frac{b}{a}\le \frac{1}{3}\frac{r_G }{r_L }\le \frac{1}{3}\quad .
\]
If $r_L \le r_G /3$, then as $r_0 \to r_G $ the sextupole to quadrupole ratio
can approach 1. In other words, for both $a$ and $b$, one must compare the
difference distributions between voids and clumps. The behavior of the tail
at large moments will have a different distinguishable behavior depending on
the radii of the lensing masses.

The interior may also be probed with the $ d $ coefficient, which
is zero except within distributions. Its strength distribution is
also shown in Fig. \ref{strength}. Over its limited range (the
maximum is at $r_0 =r_L )$ its value is surprisingly large. The
expression for finding $d $  from moments is given by equation
(\ref{eq14}).  Evidence suggesting that galaxy light paths are
penetrating lensing mass distributions is present in the angular
asymmetry evident in Fig. \ref{QdCOR}.  The background
distribution of d would be expected to be symmetric.

\section{Summary}

We visually examined faint images selected by the SExtractor
software from the Hubble deep fields using an unusually high
threshold. After filtering images with two or more maxima, we
measured sextupole and quadrupole moments. The ``curved'' galaxies
we sought were identified as those whose sextupole moment was
oriented so that one of its minima was aligned within a few
degrees of a quadrupole minimum. We then looked for and found an
improbably large spatial clumping in each Hubble deep field of
both curved and aligned galaxies. The probability of each is the
order of 1{\%}.

Our motivation and preferred hypothesis is that these galaxies
were lensed by close collisions with, for example, 10$^{10}$ solar
mass clumps that reside within a half dozen  groups of mass a few
times 10$^{12}$ solar mass in each field. The projected spacing of
such clumps within these groups would be about \textit{4.5 kpc}.
The rms radii of the footprint in the lensing plane of the
observed background galaxies is about \textit{0.4 kpc,} a factor
of 10 smaller than this spacing. The cores of these galaxies could
well be the order of \textit{2 kpc,} so there is a hope that the
effects of light paths traveling through clump interiors may be
observable through the $d$ coefficient introduced above.

We have carried out other tests that support the hypothesis that
the observed clumping comes from lensing: 1) aligned galaxies were
predicted and found to be clumped as strongly as curved galaxies,
2) galaxies halfway between aligned and curved were predicted and
found to have no clumping, and 3) correlations were predicted to
be more readily detected for galaxies with smaller moments.

We have constructed alternate hypotheses for our observations
based on instrumental effects, computational effects, or other
physical phenomena. The tests we have constructed and their
implications for each alternative hypothesis were discussed and
the results summarized in Table I. These alternatives can be
eliminated (or confirmed) with more data.

Finally, the numbers that we are seeing have very interesting and
plausible magnitudes, they even may be expected in the standard
structure formation scenarios \cite{Davis:rj}.

There are four important future studies:

\begin{enumerate}
\item Understand the image construction process of the deep-field
images to insure that the clumping property is not an artifact of
this process; \item Simulations of small-impact parameter lensing
to quantify the relationship between moment distributions and mass
distributions and to discern the fraction of galaxies that are
lensed but not seen; \item Improvement of the image-analysis
process, including variable thresholds and point-spread function
removal;  \item Deep field studies with the additional pictures
from the ACS camera on Hubble that will be available in the coming
months and years.
\end{enumerate}

In truth our observations should not be considered to be a subtle
effect. Our samples from each Hubble deep field contains only the
order of 350 background galaxies satisfying our selection
criteria. About 75 of these are ``curved'', and of these about 60
appear to reside in clumps. In other words, some 16{\%} of
background galaxies would be experiencing observable ``small
impact parameter'' scattering from dark matter clumps.

These results are possible because i) the background sextupole
strengths are a factor of 4 smaller than the quadrupole moments,
ii) the sextupole and quadrupole moment orientations arising from
lensing are correlated, and iii) one can look for clumping on the
sky.

The Hubble deep fields are less than 2 min by 2 min, about
$10^{-3}$ sq. deg.  Projects in planning stages (eg. SNAP) have a
weak lensing program of between 300 and 1000 sq. degrees with
resolution comparable to the Hubble deep field observations.

If our conjecture were to be true, weak sextupole lensing could
provide valuable insight into mass structure at length scales 100
times smaller than weak quadrupole lensing, at smaller scales than
the Lyman alpha forest results \cite{Tegmark:2002cy}.

\acknowledgements

We would like to especially acknowledge the encouragement and
support of Tony Tyson and David Wittman of Bell Labs. Visits by
both of us to Bell-Labs prepared us for this undertaking,
especially introducing us to existing software and techniques.

We would also like to thank Pisin Chen and Ron Ruth at SLAC for
trusting in our judgment and encouraging us to proceed. This work
was supported by DOE grant  DE-AC03-76SF00515.


\begin{thebibliography}{4}

\bibitem{supernova}
S.~Perlmutter {\it et al.}, ``Measurements of Omega and Lambda
from 42 High-Redshift Supernovae,'' Astrophys.\ J.\  {\bf 517},
565 (1999) [astro-ph/9812133], see also http://snap.lbl.gov;
%\cite{Riess:1998cb}
\bibitem{Riess:1998cb}
A.~G.~Riess {\it et al.}  [Supernova Search Team Collaboration],
``Observational Evidence from Supernovae for an Accelerating
Universe and a Cosmological Constant,'' Astron.\ J.\  {\bf 116},
1009 (1998) [arXiv:astro-ph/9805201].
%%CITATION = ASTRO-PH 9805201;%%

%\cite{Tonry:2003zg}
\bibitem{Tonry:2003zg}
J.~L.~Tonry {\it et al.},
%``Cosmological Results from High-z Supernovae,''
arXiv:astro-ph/0305008.
%%CITATION = ASTRO-PH 0305008;%%


%\cite{Bond}
\bibitem{Bond} J. L. Sievers {\it et al.}, ``Cosmological
Parameters from Cosmic Background Imager Observations and
Comparisons with BOOMERANG, DASI, and MAXIMA,'' astro-ph/0205387;
J.~R.~Bond {\it et al.}, ``The cosmic microwave background and
inflation, then and now,'' arXiv:astro-ph/0210007.
%%CITATION = ASTRO-PH 0210007;%%

%\cite{Spergel:2003cb}
\bibitem{Spergel:2003cb}
D.~N.~Spergel {\it et al.}, ``First Year Wilkinson Microwave
Anisotropy Probe (WMAP) Observations: Determination of
Cosmological Parameters,'' arXiv:astro-ph/0302209.
%%CITATION = ASTRO-PH 0302209;%%

%\cite{Bartelmann:1999yn}
\bibitem{Bartelmann:1999yn}
M.~Bartelmann and P.~Schneider, Phys.\ Rep.\ {\bf 340 }, 291,
(2001), arXiv:astro-ph/9912508.
%%CITATION = ASTRO-PH 9912508;%%

%\cite{Mellier:1998pk}
\bibitem{Mellier:1998pk}
Y.~Mellier,
%``Probing the Universe with Weak Lensing,''
arXiv:astro-ph/9812172.
%%CITATION = ASTRO-PH 9812172;%%

%\cite{Hoekstra:2002nf}
\bibitem{Hoekstra:2002nf}
H.~Hoekstra, H.~Yee and M.~Gladders,
%``Current status of weak gravitational lensing,''
New Astron.\ Rev.\  {\bf 46}, 767 (2002) [arXiv:astro-ph/0205205].
%%CITATION = ASTRO-PH 0205205;%%




 %% large scale mesurments
%\cite{Kaiser:2000if}
\bibitem{Kaiser:2000if}
N.~Kaiser, G.~Wilson and G.~A.~Luppino,
%``Large-Scale Cosmic Shear Measurements,''
arXiv:astro-ph/0003338.
%%CITATION = ASTRO-PH 0003338;%%

%\cite{Wittman:2000tc}
\bibitem{Wittman:2000tc}
D.~M.~Wittman, J.~A.~Tyson, D.~Kirkman, I.~Dell'Antonio and
G.~Bernstein,
%``Detection of weak gravitational lensing distortions of distant
% galaxies by cosmic dark matter at large scales,''
Nature {\bf 405}, 143 (2000) [arXiv:astro-ph/0003014].
%%CITATION = ASTRO-PH 0003014;%%

 %\cite{Mellier:2002vp}
\bibitem{Mellier:2002vp}
Y.~Mellier, L.~van Waerbeke, E.~Bertin, I.~Tereno and
F.~Bernardeau,
%``Wide-field cosmic shear surveys,''
arXiv:astro-ph/0210091.
%%CITATION = ASTRO-PH 0210091;%%

%\cite{Peebles:xt}
\bibitem{Peebles:xt}
P.~J.~Peebles, ``Principles Of Physical Cosmology,'' Princeton,
USA: Univ. Pr. (1994).
%\href{http://www.slac.stanford.edu/spires/find/hep/www?irn=2994666}{SPIRES entry}


%\cite{Huterer:2000mj}
\bibitem{Huterer:2000mj}
D.~Huterer and M.~S.~Turner,
%``Probing the dark energy: Methods and strategies,''
Phys.\ Rev.\ D {\bf 64}, 123527 (2001) [arXiv:astro-ph/0012510].
%%CITATION = ASTRO-PH 0012510;%%

%\cite{Huterer:2001yu}
\bibitem{Huterer:2001yu}
D.~Huterer,
%``Weak Lensing and Dark Energy,''
Phys.\ Rev.\ D {\bf 65}, 063001 (2002) [arXiv:astro-ph/0106399].
%%CITATION = ASTRO-PH 0106399;%%



%\cite{Linder:2003dr}
\bibitem{Linder:2003dr}
E.~V.~Linder and A.~Jenkins,
%``Cosmic Structure and Dark Energy,''
arXiv:astro-ph/0305286.
%%CITATION = ASTRO-PH 0305286;%%

%\cite{Linde:2002gj}
\bibitem{Linde:2002gj}
A.~Linde,
%``Inflation, quantum cosmology and the anthropic principle,''
arXiv:hep-th/0211048.
%%CITATION = HEP-TH 0211048;%%

%\cite{Kallosh:2003mt}
\bibitem{Kallosh:2003mt}
R.~Kallosh and A.~Linde,
%``Dark energy and the fate of the universe,''
JCAP {\bf 0302}, 002 (2003) [arXiv:astro-ph/0301087].
%%CITATION = ASTRO-PH 0301087;%%

%\cite{Kallosh:2003bq}
\bibitem{Kallosh:2003bq}
R.~Kallosh, J.~Kratochvil, A.~Linde, E.~V.~Linder and M.~Shmakova,
%``Observational Bounds on Cosmic Doomsday,''
arXiv:astro-ph/0307185.
%%CITATION = ASTRO-PH 0307185;%%


%%%%%% SNAP REfer %%%%%%%%%%%%%%%%%%
%\cite{unknown:2002dp}
\bibitem{unknown:2002dp}
  G Aldering [the SNAP Collaboration],
%``Overview of the SuperNova / Acceleration Probe (SNAP),''
arXiv:astro-ph/0209550.
%%CITATION = ASTRO-PH 0209550;%%

%\cite{SNAP}
\bibitem{SNAP}
J.~Rhodes, A.~Refregier and R.~Massey  [the SNAP Collaboration],
%``Weak Lensing from Space I: Prospects for The Supernova/Acceleration Probe,''
arXiv:astro-ph/0304417;
%%CITATION = ASTRO-PH 0304417;%%
R.~Massey {\it et al.},
%``Weak Lensing from Space II: Dark Matter Mapping,''
arXiv:astro-ph/0304418;
%%CITATION = ASTRO-PH 0304418;%%
A.~Refregier {\it et al.},
%``Weak Lensing from Space III: Cosmological Parameters,''
arXiv:astro-ph/0304419.
%%CITATION = ASTRO-PH 0304419;%%

%%%%%%%%%%%%%%% LSST %%%%%%%%%%%%%%
%\cite{Tyson:2003kb}
\bibitem{Tyson:2003kb}
J.~A.~Tyson  [the LSST Collaboration],
%``Large Synoptic Survey Telescope: Overview,''
Proc.\ SPIE Int.\ Soc.\ Opt.\ Eng.\  {\bf 4836}, 10 (2002)
[arXiv:astro-ph/0302102].
%%CITATION = ASTRO-PH 0302102;%%

 %\cite{finalfocus}
\bibitem{finalfocus}
J.~ Irwin and M.~Shmakova, ``Cosmological Final Focus Systems'',
to be published in the proceedings of 28-th Workshop on ``Quantum
Aspects of Beam Physics.''

%\cite{Williams:1996ay}
\bibitem{Williams:1996ay}
R.~E.~Williams  [the HDF Team Collaboration],
%``The Hubble Deep Field: Observations, Data Reduction, and Galaxy Photometry,''
arXiv:astro-ph/9607174.
%%CITATION = ASTRO-PH 9607174;%%

%\cite{Casertano:2000wy}
\bibitem{Casertano:2000wy}
S.~Casertano {\it et al.},
%``WFPC2 Observations of the Hubble Deep Field-South,''
arXiv:astro-ph/0010245.
%%CITATION = ASTRO-PH 0010245;%%

%\cite{sextractor}
\bibitem{sextractor}
E.~Bertin and S.~Arnouts, Astron. Astrophys. Suppl. Ser.{\bf 117},
393 (1996)


%%%%% large scale structure Check !!!!
%\cite{Davis:rj}
\bibitem{Davis:rj}
M.~Davis, G.~Efstathiou, C.~S.~Frenk and S.~D.~White,
%``The Evolution Of Large-Scale Structure In A Universe Dominated By Cold  Dark Matter,''
Astrophys.\ J.\  {\bf 292}, 371 (1985).
%%CITATION = ASJOA,292,371;%%

%\cite{Tegmark:2002cy}
\bibitem{Tegmark:2002cy}
M.~Tegmark and M.~Zaldarriaga,
%``Separating the Early Universe from the Late Universe: cosmological parameter estimation beyond the black box,''
Phys.\ Rev.\ D {\bf 66}, 103508 (2002) [arXiv:astro-ph/0207047].
%%CITATION = ASTRO-PH 0207047;%%


\end{thebibliography}
\end{document}